\documentclass[12pt,preprint]{aastex}
\usepackage{natbib,amssymb,amsmath,graphicx}
\newcommand\tna{\,\tablenotemark{a}}
\newcommand\tnb{\,\tablenotemark{b}}
\newcommand\tnc{\,\tablenotemark{c}}
\newcommand\tnd{\,\tablenotemark{d}}
\newcommand\tne{\,\tablenotemark{e}}
\newcommand\tnf{\,\tablenotemark{f}}
\newcommand\tng{\,\tablenotemark{g}}
\newcommand\tnh{\,\tablenotemark{h}}
\newcommand\tni{\,\tablenotemark{i}}
\newcommand\tnj{\,\tablenotemark{j}}
\begin{document} 
\title{Disentangling Confused Stars at the Galactic Center with Long Baseline Infrared Interferometry}
\author{Jordan M. Stone, J.~A. Eisner}

\affil{Steward Observatory, 
University of Arizona,
933 N. Cherry Ave,
Tucson, AZ 85721-0065, USA;
jstone,jeisner@email.arizona.edu}

\author{J.~D. Monnier}

\affil{Astronomy Department,
University of Michigan,
941 Dennison Bldg,
Ann Arbor, MI 48109-1090, USA}

\author{J. Woillez, P. Wizinowich}

\affil{W.M. Keck Observatory,
65-1120 Mamalahoa Highway,
Kamuela, HI 96743, USA}

\author{J.-U. Pott}

\affil{Max-Planck-Institut f\"{u}r Astronomie,
K\"{o}nigstuhl 1,
 D-69117 Heidelberg, Germany}

\author{A.~M. Ghez}

\affil{Department of Physics and Astronomy, 
UCLA, 
Los Angeles,CA 90095-1547, USA}

\begin{abstract} 
We present simulations of Keck Interferometer ASTRA and VLTI
GRAVITY observations of mock star fields in orbit within $\sim50$
milliarcseconds of Sgr A*. Dual-field phase referencing techniques, as
implemented on ASTRA and planned for GRAVITY, will provide the sensitivity to
observe Sgr A* with long-baseline infrared interferometers. Our results show an
improvement in the confusion noise limit over current astrometric surveys,
opening a window to study stellar sources in the region.  Since the Keck
Interferometer has only a single baseline, the improvement in the confusion
limit depends on source position angles. The GRAVITY instrument will yield a more compact and symmetric PSF,
providing an improvement in confusion noise which will not depend as strongly
on position angle. Our Keck results show the ability to characterize the star
field as containing zero, few, or many bright stellar sources. We are also able
to detect and track a source down to $m_K\sim18$ through the least confused
regions of our field of view at a precision of $\sim200~\mu$as along the
baseline direction. This level of precision improves with source brightness.
Our GRAVITY results show the potential to detect and track multiple sources in
the field. GRAVITY will perform $\sim10~\mu$as astrometry on a $m_K=16.3$
source and $\sim200~\mu$as astrometry on a $m_K=18.8$ source in six hours of
monitoring a crowded field. Monitoring the orbits of several stars will provide
the ability to distinguish between multiple post-Newtonian orbital effects,
including those due to an extended mass distribution around Sgr A* and to
low-order General Relativistic effects. ASTRA and GRAVITY both have the
potential to detect and monitor sources very close to Sgr A*. Early
characterizations of the field by ASTRA including the possibility of a precise
source detection, could provide valuable information for future GRAVITY
implementation and observation.
\end{abstract}

\section{Introduction}
Over the last 20 years, high resolution infrared imaging techniques have
provided precise astrometric measurements of stellar sources at the Galactic
Center. Focused astrometric monitoring campaigns have revealed a population of
mostly young early-type stars (the S-cluster) in orbit about the location of
the radio and infrared source dubbed Sagittarius A* (Sgr A*). In fact,
\citet{Ghez08} and \citet{Gill09a} both analyze their own distinct data sets to
deduce a mass of $4.5\pm0.4\times10^{6} \mathrm{M}_{\odot}$ located coincident
with Sgr A*. This mass must all be within the periapsis of the star
S16/S0-16\footnote{The UCLA and the MPE groups have adopted different naming
conventions for the S-cluster}, which is only ~40 AU. The implied mass density
provides compelling proof that Sgr A* is the luminous manifestation of an
accreting black hole. In addition to providing a measurement of the mass of Sgr
A*, the orbits of the stars also provide a direct measurement of the distance
to the black hole, $8.36\pm0.44$ kpc \citep{Ghez08}.

If Sgr A* actually resides at the dynamic center of the Milky Way, then
measuring its distance also represents a measurement of the solar distance from
the Galactic Center ($R_0$). Monitoring stellar orbits about Sgr A* also
provide a measurement of the sun's peculiar motion in the direction of the
Galactic Center ($\Theta_0$).  As discussed by \citet{Olling00}, $R_0$ and
$\Theta_0$ are ubiquitous parameters in the description of the structure and
dynamics of the Milky Way \citep[see also][]{Reid2009}.  Uncertainty in the
values of $R_0$ and $\Theta_0$ are the largest sources of error in the
determination of the ratio of the galactic halo's long and short axes
\citep[$q$,][]{Olling00}.  The parameter $q$ is sensitive to different galaxy
formation scenarios and dark matter candidates, and if $R_0$ and $\Theta_0$
were known at the $1\%$ level, the constraints on $q$ would help to
differentiate theories of galaxy formation and dark matter \citep{Olling00}.
In addition, a very precise knowledge of $R_0$ could affect our calibration of
the lowest rungs of the cosmic distance ladder by improving our distance
measurements to galactic sources such as Cepheids and RR Lyrae
variables\citep{Ghez08}.

The existence of the S-cluster is intriguing because it is rich with young
stars and because forming these stars in situ represents a theoretical problem
given the strong tidal forces in the region. The alternative of formation at
larger radii and subsequent migration is strongly constrained by the deduced
young age of the stars.  However, because the S-stars are known to be younger
than the relaxation time in the environment \citep[a B star's main-sequence lifetime
is $\sim10^{7}$ years compared to the relaxation time of $\sim 2\times10^8$
years;][]{Weinberg05} their orbits should encode some information about the
kinematics of the cluster at the time it formed. Thus, perhaps as a bonus,
astrometric monitoring of the stars in our Galaxy's nuclear cluster has the
potential to inform the community not only on matters of General Relativity and
galaxy formation, but also on star formation in extreme environments.

The deduced mass and distance of Sgr A* make it the largest black hole on the
sky, in terms of angular diameter, and an excellent candidate for study.
Improving astrometric measurements and discovering stars on even shorter period
orbits will improve our understanding of the gravitational potential which
binds the stars, possibly exposing a dark matter cusp at the center of our
galaxy, and should inform our understanding of gravity on scales not yet
explored by precise experiments. For example, \citet{Weinberg05} modeled
a distribution of stars on very short period orbits about Sgr A* and showed
that post-Newtonian effects on the orbital paths could be detected with the
astrometric precision and sensitivity of a future thirty-meter telescope.
Existing and upcoming near-IR interferometers can provide even better
resolution, and enable some of the same science. In their treatment
\citet{Weinberg05} assumed Gaussian point spread functions, which tend to zero
in the wings faster than the more realistic Airy pattern which distributes
light in rings away from the central core. These rings present a contrast
barrier in conventional imaging. Likewise, the even more complicated point
spread functions (PSFs) provided by interferometers result in contrast and
detection limits and can bias astrometric measurements. 

\citet{Fritz10} showed that halo noise and source confusion are the factors
limiting astrometric accuracy. These effects, which are both related to the
angular resolution of the telescope and the luminosity function of the sources
in the region (i.e. the dynamic range), present a fundamental astrometric
hurdle which cannot be overcome with even the largest single aperture
telescopes of today \citep{Ghez08,Gill09a}.  In fact, simulations by
\citet{Ghez08} and \citet{Gill09a} showed that astrometric errors could be as
large as 3 milliarcseconds due to confusion with undetected sources. This level
of astrometric uncertainty, present close to Sgr A* where stars
experience the deepest part of the potential, has precluded the detection of
any post-Newtonian effects on stellar orbits to this point and has limited the
precision with which the distance to the Galactic Center can be measured.  

The limiting magnitude and astrometric precision of Galactic Center
observations has improved as early speckle observations
\citep[e.g.][]{Eckart92} were supplemented by adaptive optics (AO) and laser
guide star AO \citep[e.g.][]{Ghez05}. The current state of the art is
a limiting magnitude of $\sim$19 at K and an astrometric precision of
$\sim100\mu$as.  However, these limits are only achieved far from the crowded
central region immediately surrounding Sgr A*. In this paper we investigate
whether, with the increased resolution of infrared interferometers and the
concomitant reduction in confusion, we can detect heretofore undetected or
unnoticed stars on orbits with very short periods within 50 mas of Sgr A*. We
also explore the astrometric precision with which such sources could be
monitored with an IR interferometer. According to \citet{GhezTalk} a factor of
$\sim3$ more stars with periods less than 20 yrs are expected to be orbiting
Sgr A*. These stars, if they can be detected and monitored, will provide
a detailed description of the central potential 
\citep[a minimum of three short-period orbits are required for a complete characterization;][]{Rubilar2001}. Additionally, such stars will provide the best targets for
observing General Relativistic effects since they are deepest in the potential
well of Sgr A*. 

Although the higher resolution provided by interferometry is beneficial for
increasing the detectability of sources in the crowded region and for
increasing the astrometric precision attainable, there are many potential
complicating factors which do not apply to conventional full aperture imaging.
For example, in full aperture imaging, collecting area increases as the square
of the resolution.  In interferometers, however, the collecting area is
independent of the effective spatial resolution. This means that although the
confusion limit is somewhat alleviated by the higher angular resolution
available, photon noise quickly becomes a problem in the detection of faint
sources.  This fact is further exacerbated by the low typical throughput of
interferometers (e.g., $\sim2\%$ for the Keck Interferometer).  Additionally,
the sparse nature of an interferometer's collecting area results in an
incomplete sampling of the Fourier components of the source distribution on the
sky.  This causes an incomplete knowledge of the sky-plane light distribution
resulting in PSFs with large sidelobes.  Finally, Michelson interferometers
like the Keck Interferometer and the VLT Interferometer (VLTI) have small fields of
view, $\sim50$ mas, which typically only include a single object; clearly this
presents difficulty for astrometry.  We attempt to understand the scale of
these effects by simulating data and inferring results.

The outline of the paper is as follows.  In Section 2 we discuss the
construction of mock star fields within 50 mas of the Galactic Center.  In
Section 3 we discuss our observation simulator and all included sources of
noise and uncertainty. Section 4 covers our algorithm for making relative astrometric
measurements by fitting to the visibility curves.  Section 5 includes
a presentation of our results.  Section 6 provides a discussion of the
potential advances and difficulties. Although we try to keep the discussion
general, we focus on ASTRA at the Keck Interferometer as we are most familiar
with that instrument and it is currently capable of making these observations.
We also provide a discussion of VLTI GRAVITY \citep{GRAVITY3} simulations. 

\section{Simulating a Star Field within 50 Milliarcseconds of Sgr A*} 

Interferometric observations of the Galactic Center will be conducted through
the K band.  We construct mock star fields in orbit about the black hole at the
Galactic Center using as much information about the number of sources and the
distribution of K band magnitudes as is available. Previous observations, while limited
by confusion noise, have provided a wealth of information regarding the stellar
population and K-band luminosity function (KLF) of the central cluster.
\citet{Genzel03} showed that the KLF of the inner 1.5" around Sgr A* is well
fit by a power law with a slope of
$\beta=\frac{d\mathrm{log}N}{dK}=0.21\pm0.03$.  \citet{Weinberg05}, for example,
used this luminosity function normalized to the photometry of \citet{Schodel02}
of the stars within 0.8" of Sgr A* to extrapolate the population of stars even
closer to the Galactic Center.  

Additionally, some clues about the content of the confused region can be gleaned from
observations of well-monitored stars--- stars on orbits which spend most of
their time outside of the confused region--- when they pass through the
confused region during their orbital periapsis.  For example, when the star
S0-2 passed through periapsis in 2002 its centroid was offset from its fitted
orbital path. This offset strongly suggests a confusion event with an
undetected source (or multiple undetected objects).  By measuring the offset
and the ellipticity, or lack thereof, of the S0-2 PSF during 2002, some
constraints on the unseen source(s) can be derived \citep[e.g.][]{Gill09a}. 

Further, both \citet{Do09} and \citet{Dodds-Eden11} provide analyses of Sgr A*
light curves and show a median magnitude in K-band of $\gtrsim$ 16 and
a minimum magnitude of $\gtrsim$ 17. While absolute photometry in the region is
complicated by confusion noise, we use the analyses of Sgr A* lightcurves to
normalize two separate star fields for use in our study. In our first star
field, Field1, we set the flux from Sgr A* equal to its median observed value,
the flux from the brightest star in the field we set to be consistent with the
minimum observed flux from the region as reported in \citet{Do09} and
\citet{Dodds-Eden11}, and we include three fainter stellar sources consistent
with an extrapolation of the KLF reported in \citet{Genzel03} (see their Table
2). To be cautious, we also model a second star field in which we have
decreased the flux of each source by one magnitude. This fainter star field we
call Field2. We believe that these two fields should encompass a fair
representation of the true source distribution very close to Sgr A*.

While we took care that our star fields are consistent with available
observations, the true source content within 50 milliarcseconds is unknown. The
potential to further constrain the stellar content within the region is one the
scientific motivations for observing the Galactic Center with long-baseline
infrared interferometers.

We assigned spectral slopes to our sources to be
consistent with the observed slopes of sources near the Galactic Center, namely
$\alpha=2.3\pm0.1$ for $F_{\lambda} \propto \lambda^{\alpha}$.  The spectral
slope of Sgr A* was taken to be $\alpha=4.5$ (Tuan Do, private communication;
see also \citet{Do09B} and \citet{Paumard06}).

While we modeled Sgr A* as stationary, the stellar sources were assigned random
Keplerian orbits with semi-major axes in the range of 0.1-400 AU. We chose this
range of semi-major axes to coincide with the confused region around Sgr A*.
Any source with a larger semi-major axis should leave this region, and would
likely have been detected in previous AO observations. Table \ref{orbitTable}
shows the modeled parameters for our star fields.
\begin{deluxetable}{cccccccccc}
\tabletypesize{\scriptsize}
\tablecolumns{9}
\tablewidth{0pt}
\tablecaption{Model star field parameters for our simulated fields.}
\tablehead{
    \colhead{Source} &
    \colhead{Field1 [$m_K$]} & 
    \colhead{Field2 [$m_K$]} & 
    \colhead{spectral slope} & 
    \colhead{Period [yrs]} & 
    \colhead{e} & 
    \colhead{$\tau [\mathrm{yrs}]$} & 
    \colhead{$\Omega [\mathrm{rad}]$} & 
    \colhead{$i [\mathrm{rad}]$} & 
    \colhead{$\omega [\mathrm{rad}]$} }       
\startdata
Sgr A*   & 16.3 & 17.3 & 4.5 &--&--&--&--&--&--\\ 
Star 1   & 16.9 & 17.9 & 2.28 & 1.75 & 0.44 & 0.93 & 5.32 &  0.68 & 4.19\\ 
Star 2   & 18.8 & 19.8 & 2.34 & 2.65 & 0.12 & 0.60 & 2.61 &  1.36 & 4.89\\ 
Star 3   & 20.3 & 21.3 & 2.31 & 0.24 & 0.90 & 0.20 & 4.64 & -0.61 & 2.95\\ 
Star 4   & 20.6 & 21.6 & 2.14 & 2.31 & 0.74 & 1.53 & 1.16 &  1.07 & 0.91\\ 
\enddata
\tablecomments{
We assumed random Keplerian orbits for our modeled sources.  $\tau$ is the time
of periapsis passage in years since January 1, $\Omega$ is the longitude of the
ascending node, $i$ is the inclination, and $\omega$ is the argument of
periapsis. \label{orbitTable}}
\end{deluxetable}

\section{Synthesizing Visibility Data}\label{simulateSec}
\subsection{Basic Concepts}

Interferometers combine light from a source collected at more than one
aperture, forming an interference pattern.  This interference pattern encodes
high spatial frequency information of the light distribution under observation.
The maximum spatial frequency which can be detected by an interferometer can be
expressed succinctly as $\frac{B_{\mathrm{proj}}}{\lambda}$, where $\lambda$ is
the wavelength of the observation and $B_{\mathrm{proj}}$ is defined as the
projection of the vector connecting the two apertures onto the plane of the sky
in the direction of observation. The interference pattern of light created by
a monochromatic source can be described as
\begin{equation}
P(\delta)= P_{1}+P_{2}+2\sqrt{P_{1}P_{2}} \mathrm{Re}\{V e^{ik\delta}\})
\end{equation}
or
\begin{equation}
P(\delta)= P_{1}+P_{2}+2\sqrt{P_{1}P_{2}}|V|\cos(k(\mathrm{arg}(V)+\delta)))
\end{equation}
where $P_{1}$ and $P_{2}$ are the incident power from each aperture, $k$ is the
wave number, $\delta$ is the relative path-length delay between the two
apertures, and $V$ is the complex visibility \citep[see e.g.][]{Lawson2000}. 

The polychromatic case is slightly more complicated, and a brief discussion of
the effects of a finite bandwidth on interference fringes will help elucidate
some of the challenges of practical interferometry.  In general, astronomical
sources radiate polychromatic light such that each wavelength of light is
mutually incoherent with every other wavelength.  The result of this mutual
incoherence is that polychromatic fringes are a sum of
fringes produced by each individual wavelength of light.  Since a fixed
$\delta$ will result in a different phase offset between the two apertures for
each wavelength of light (i.e. $\phi=k\delta$), it is impossible
to keep fringes arising from different wavelengths of light in phase
everywhere.  Polychromatic fringes are only present in the neighborhood of zero
delay and are attenuated as $\delta$ increases.  In fact, the fringes are
attenuated by an envelope function which takes the shape of the Fourier
transform of the spectral bandpass.  For the case of a top-hat bandpass
centered at $\lambda_0$ and with a width of $\Delta \lambda$, the envelope
takes the shape of a sinc function, and fringes are detected inside the
coherence length 
\begin{equation} 
\Lambda \equiv \frac{\lambda_{0}^{2}}{\Delta \lambda}.
\end{equation}
For a typical case of five spectral channels in the K band,
$\Lambda\sim25\lambda$ (see Table \ref{specTable}).  This imposes a strict
requirement on the implementation of interferometric measurements.  In general,
the path lengths of light incident on different telescopes must be similar to
within $\Lambda$ before being combined if fringes are to be observed at all.
Even more strictly, $\delta$ must be less than $\frac{\lambda}{B_{\mathrm{proj}}}$ if the
central unattenuated ``white-light" fringe is to be observed. This necessitates
delay lines to precisely correct for the optical path length difference between
the two apertures.  Further, the existence of a coherence envelope also imposes
a restriction on our field of view. For a Michelson Interferometer the field of
view can be calculated according to
\begin{equation}
\mathrm{FOV}=\frac{\Lambda}{B_\mathrm{proj}}.
\end{equation}
Table \ref{specTable} lists $\Lambda$ and the corresponding field of view for
each of the modeled spectral channels using the Keck Interferometer.
\begin{deluxetable}{cccc}
\tabletypesize{\scriptsize}
\tablecolumns{4}
\tablewidth{0pt}
\tablecaption{Coherence Length}
\tablehead{
    \colhead{$\lambda_0$ [microns]} & 
    \colhead{$\Delta \lambda$ [microns]} & 
    \colhead{$\Lambda$ [microns]} &
    \colhead{Field of View [mas]}  }
\startdata
2.0   &  0.087  &  45.97 & 141.99  \\ 
2.09  &  0.091  &  48.00 & 148.25  \\ 
2.18  &  0.095  &  50.02 & 154.49  \\ 
2.28  &  0.099  &  52.50 & 162.15  \\ 
2.38  &  0.104  &  54.46 & 168.21  \\ 
\enddata
\tablecomments{The central wavelength and width of the spectral channels for
our observations. Also shown are the corresponding coherence length and implied
field of view.  As we will see in Section \ref{uncSection}, the true limiting
factor for our field of view will be the fiber response function which is
modeled to have a full width at half maximum of 55 mas.\label{specTable}
} \end{deluxetable}

Atmospheric turbulence induces a differential delay between the two
telescopes, $\delta_{\mathrm{atmosphere}}$, which is variable on timescales of
$t_\mathrm{turb}\sim\frac{D}{v_{\mathrm{wind}}}$, with $D$ the diameter of each aperture
and $v_{wind}$ the wind velocity.  If an exposure time of longer than $\sim
t_{\mathrm{turb}}$ is required, the delay lines at the interferometer, which
are responsible for keeping $\delta\approx0$, cannot only compensate for sidereal motion
or else fringes will be smeared; atmospheric delay must be
detected and corrected for dynamically. 

Because the Galactic Center is faint, a dual field system is required
to incorporate a bright reference star for fringe tracking.  In dual field
phase referencing (DFPR) a field separator is used to create two beam trains,
one with a bright reference source and one with a fainter science target. Similar to 
natural guide star adaptive optics, light from the bright reference star is used
to detect and measure the differential atmospheric delay. These measurements
are enabled by the high flux of the reference source, which allows for short
exposures of less than $t_\mathrm{turb}$. 

Precise metrology is required for the application of a dual field system.
Delay line commands need to be precise on the order of a fraction of the fringe
spacing, which is typically $\sim0.03$ microns for our scenarios observing in the K band. Also
important, a precise baseline measurement is needed to perform the conversion
from delay offset of the reference fringes to a delay line command in the
science beam.  The situation is complicated since there are two beamtrains,
each with distinct path lengths which must be measured continuously to monitor
fluctuations due to thermal drifts and other effects. 

\subsection{Sources of Uncertainty for Phase Referenced Interferometric Imaging}\label{uncSection}

Visibility measurements entail measuring the amplitude and phase of
interference fringes using a light detector. Any corruption of fringe
amplitude, fringe phase, or flux levels on the detector must be taken into
account. Below we describe all such influences considered.

A fringe-tracker measures the phase of the fringes in the reference beam and
sends corrections to delay lines.  The finite servo bandwidth in the
fringe-tracker will cause the observed fringes to be smeared, and lowering the
observed amplitude by a factor which we take to be:
\begin{equation}
C_{\mathrm{servo}}=0.75, 
\end{equation} 
consistent with the performance of phase referencing at the Palomar Testbed
Interferometer \citep{Lane2003}.

In DFPR the reference source and the science target are observed through
different portions of the atmosphere, each with slightly different piston
aberrations \citep[anisopistonism,][]{Esposito00, Colavita09}. Thus the
assumption that light from the science target can be held in phase by
monitoring the phase of the brighter reference source will introduce an error.
This error will smear the fringes and reduce the modulus of the visibility.
The size of this effect is discussed in \citet{Esposito00} and
\citet{Colavita09} and is given as 
\begin{equation}\label{anisoEq}
C_{\mathrm{aniso}}=\mathrm{exp}(-0.44(\frac{\theta}{\theta_p})^2)
\end{equation}
where $\theta_p$ is the isopistonic angle given by \citet{Esposito00}, and
$\theta$ is the angle to the reference star.  

After the cophased beams are combined, the light passes through single-mode
fiber optic cables and then to the camera. Fibers respond best to on-axis
sources and transmit only a fraction of the light from off-axis sources.  This
reduces the observed flux of each source by a factor of $F(\alpha,\beta)$,
where $F$ is the fiber function defined below and $\alpha$ and $\beta$ are the
position of each source relative to the phase center.  We model the
fiber response function as achromatic with a Gaussian function having a full
width at half maximum of 55 milliarcseconds:
\begin{equation}\label{FiberEQ}
F(\alpha,\beta)=\exp(-4\log(2)\frac{\alpha^2+\beta^2}{(55 \mathrm{mas})^{2}}).
\end{equation}

Pointing errors also affect our astrometric precision. In interferometry,
pointing errors are manifest as phase errors in the complex visibilities.  We
account for phase errors arising from a handful of instrumental effects
primarily related to measuring and monitoring the baseline and internal paths.
We label the combined contribution of these effects
$\sigma_{\mathrm{metrology}}$.
Another phase error that affects our ability to point arises due to the
different paths light from separate sources take through the atmosphere.  As
discussed by \citet{Shao92}, the uncertainty in the observed phase of
interferometric fringes can be written as
\begin{equation}
\sigma_{\mathrm{atmosphere}} \approx 300B^{\frac{-2}{3}}\theta t^{\frac{-1}{2}}~\mathrm{arcseconds},\label{sigAtmos} 
\end{equation}
where $B$ is the projected baseline in meters, $\theta$ is the angular
separation between the science target and a reference source in radians, and
$t$ is the integration time in seconds. This relationship is calibrated using
models of the atmosphere above Mauna Kea.  We combine the two sources of phase
error to produce a total pointing error:
\begin{equation}\label{pointErr}
\sigma_{\mathrm{point}}^2=\sigma_{\mathrm{metrology}}^2+\sigma_{\mathrm{atmosphere}}^2.
\end{equation}

We measure the visibility amplitude and phase using the four-bin ABCD algorithm
\citep{Colavita99}.  In the ABCD algorithm the average intensity in each
quarter fringe (A,B,C,and D) is measured and used to deduce the complex
visibility. The real and imaginary parts of the visibility can be calculated
according to:
\begin{equation}
\mathrm{Re}\{V\}=A-C,
\end{equation}
and
\begin{equation}
\mathrm{Imag}\{V\}=B-D.
\end{equation}

Since each measurement A through D is a flux measurement, each is susceptible
to normal photometric uncertainties including Strehl fluctuations, photon
noise, and readout noise. We assume $\sigma_{\mathrm{Strehl}}=10\%$ injection
fluctuations, which account for Strehl variations and for the variable coupling
of speckles from bright sources beyond our field of view into the fiber
(speckle coupling). We assume an average Strehl in the K band of 35\%. While
this value of the Strehl is consistent with typical values in previous laser
guide star AO observations \citep[e.g][]{Ghez05}, it may represent an
optimistic estimate for performance with ASTRA. This is because reported Strehl ratios in the literature
are likely to be biased high by frame selection. In our simulations, a reduced
Strehl ratio will result in a lower level of flux in the fringes. In our
results section below, we demonstrate how our performance scales with flux by
modeling two starfields which are identical except that in one all source
fluxes have been scaled by one magnitude. 

Bright stars on the periphery of our field of view also have the potential to
affect our simulated visibilities. This is because, as shown in Table
\ref{specTable}, the interferometric coherence envelope extends out to a radius
$\sim70$ mas and because the fiber will have at least some response there. We
assume that such bright sources at large radii have already been detected with
laser guide star AO observations and that their effect on our visibilities
could be modeled out accordingly.

We report our modeled background flux levels in each spectral channel in
Table \ref{VisParams}. They are based on observed background performance using
the Keck Interferometer fringe tracker \citep[e.g.][]{Woillez12}.  The read
noise, $\sigma_{\mathrm{rdnz}}$, is taken to be 10 counts, consistent with the
performance of the Hawaii arrays at Keck.  Thus we take the uncertainty in each
read in each spectral channel to be 
\begin{equation}\label{detect}
\sigma_{\mathrm{detect}}^2
= t_{\mathrm{int}}(\sum_{j}(F(\alpha_j,\beta_j)(0.35+\eta_{1}+0.35+\eta_{2})(P_j))+P_{\mathrm{bg}}+\sigma_{\mathrm{rdnz}}^2
\end{equation} 
where $j$ indexes each source in the field, $\alpha_j$ and
$\beta_j$ are he position of each source on the fiber, $P_j$ is the flux on
each aperture emanating from each source, and $P_{\mathrm{bg}}$ is the
background flux. The exposure time $t_{\mathrm{int}}$ is the time spent on each
read, and $\eta_1$ and $\eta_2$ are sampled values of the injection
fluctuations drawn from a Gaussian distribution of standard deviation
$\sigma_{\mathrm{Strehl}}$.

Our modeled observational setup uses 60 five-second subreads per 300-second
block. Each subread includes one second of integration on each fringe
quadrature and one second on a bias frame. A summary of our simulator
parameters can be found in Table \ref{VisParams}.

\subsection{Producing Mock Fringes\label{dataSection}}
In the previous section, we described the magnitude of several corrupting influences
including several random noise sources. Actual values for each of these
random sources of noise must be realized before we can produce our simulated
fringes. We assign to the variable $\phi_{\mathrm{err}}$ a sampled value of the
random pointing error drawn from a Gaussian distribution of width
$\sigma_{\mathrm{point}}$. For each aperture we draw a value for the random
injection fluctuation from a Gaussian of width $\sigma_{\mathrm{Strehl}}$; we
assign these values to the variable $\eta_i$, where $i$ indexes the aperture.
Finally, we generate a value, $n_{\mathrm{detect}}$, for the detection noise by drawing from a Gaussian
distribution of width $\sigma_{\mathrm{detect}}$.

Next, we combine the random noise sources with calculated values for $C_{\mathrm{aniso}}$,
the fiber attenuation function, and the visibility of each individual point
source to model and measure the fringes as follows
\begin{equation}
\hat{E}(\delta)=t_{\mathrm{int}}((0.5+\eta_1)P+(0.5+\eta_2)P+
2\sqrt{(0.5+\eta_1)(0.5+\eta_{2})P^{2}}|\hat{V}|\cos(k(\mathrm{arg}(\hat{V})+\delta)))
\end{equation}
where the amplitude and phase of the fringes are specified by the complex
visibility given by
\begin{equation}
\hat{V}=\frac{\sum_{j}F(\alpha_j,\beta_j)P_{j}C_{\mathrm{servo}}C_{\mathrm{aniso}} 
V_{\mathrm{point}}(\alpha_j,\beta_j,u,v)e^{2\pi i \phi_{\mathrm{err}}}}
{\sum_{j}F(\alpha_j,\beta_j)P_{j}}.
\end{equation}
Here $V_{\mathrm{point}}(\alpha,\beta,u,v)$ is the complex visibility of a point source
\citep[see e.g.][]{Lawson2000}, $t_{\mathrm{int}}$ is the exposure time (1 second in our
model). The quadratures are then determined using 
\begin{equation}
\hat{A}=\frac{\int_{0}^{\frac{\pi}{2}}\hat{E}_{A}(\delta)\,\mathrm{d}\delta}{\frac{\pi}{2}}
+n_{\mathrm{detectA}}
\end{equation}
\begin{equation}
\hat{B}=\frac{\int_{\frac{\pi}{2}}^{\pi}\hat{E}_{B}(\delta)\,\mathrm{d}\delta}{\frac{\pi}{2}}
+n_{\mathrm{detectB}}
\end{equation}
\begin{equation}
\hat{C}=\frac{\int_{\pi}^{\frac{3\pi}{2}}\hat{E}_{C}(\delta)\,\mathrm{d}\delta}{\frac{\pi}{2}}
+n_{\mathrm{detectC}}
\end{equation}
\begin{equation}
\hat{D}=\frac{\int_{\frac{3\pi}{2}}^{2\pi}\hat{E}_{D}(\delta)\,\mathrm{d}\delta}{\frac{\pi}{2}}
+n_{\mathrm{detectD}}
\end{equation}
For each hour angle and wavelength observed, the real
and imaginary parts of the visibility are then reported as $\hat{A}-\hat{C}$
and $\hat{B}-\hat{D}$ respectively. It is important to note that a new
realization of $\hat{E}$ is made for each quadrature, exposing each quadrature
to independent injection fluctuations. This inter-ABCD fluctuation affects both
the deduced amplitude and phase of the complex visibility, and is often the
dominant noise source. 
\begin{deluxetable}{llll}
\tabletypesize{\scriptsize}
\tablewidth{0pt}
\tablecaption{Parameters}
\tablehead{
    \colhead{Parameter} &
    \colhead{Keck} & 
    \colhead{VLTI} & 
    \colhead{Shared}  
}
\startdata
Number of baselines                                      & 1                        & 6                            & --- \\
Time on source each observing night                      & 3 hours                  & 6 hours                      & --- \\
Blocktime                                                & ---                      & ---                          & 300 seconds \\
Transmission\tna                                         & 1.7\%                    & 0.9\%                        & --- \\
Fiber full width half maximum                            & ---                      & ---                          & 55 milliarcseconds \\
Strehl                                                   & ---                      & ---                          & 35\%\tnb \\
Injection Fluctuations ($\eta$)                          & ---                      & ---                          & 10\% \\
Background flux rate in each channel                     & ---                      & ---                          & 60, 167, 430, 1112, and 2677 photons/second\tnc \\
$\sigma_{\mathrm{readnoise}}$                            & ---                      & ---                          & 10 counts \tnd \\
Decoherence Due to the servo ($C_{\mathrm{servo}}$)      & ---                      & ---                          & 0.75\tne \\
Isopistonic angle ($\theta_{\mathrm{p}}$)\tnf            & 13.5 arcseconds          & 16.1 arcseconds              & ---  \\
Distance to fringe tracking star ($\theta$)\tng          & 7 arcseconds             & 1.2 arcseconds               & --- \\
Decoherence due to anisopistonism ($C_{\mathrm{aniso}}$) & 0.90                     & 0.998                        & --- \\
$\sigma_{\mathrm{atmosphere}}$\tnh                       & $32\--42 \mu \mathrm{as}$& $4\--7\mu\mathrm{as}$        & --- \\
$\sigma_{\mathrm{metrology}}$                            & $20~\mu \mathrm{as}$\tni & $14.5~\mu \mathrm{as}$\tnj       & --- \\
\enddata

\tablenotetext{a}{Demonstrated at Keck \citet{Woillez12}; GRAVITY reference: \citet{Vincent11}}
\tablenotetext{b}{Consistent with AO performancd at Keck
\citep[e.g.][]{Ghez05}, and the expected performance for the GRAVITY instrument
\citep{GRAVITY}}
\tablenotetext{c}{Background flux rates refer to the flux in the spectral
channels centered at 2.0, 2.09, 2.18, 2.28, and 2.38 microns respectively and are consistent with the observed performance on the Keck Interferometer \citep{Woillez12}}
\tablenotetext{d}{consistent with the performance of the HAWAII IIRG arrays \citep[e.g.][]{Woillez12}}
\tablenotetext{e}{From \citet{Lane2003}}
\tablenotetext{f}{From \citet{Esposito00}}
\tablenotetext{g}{IRS 7 at Keck, and IRS 16C at the VLTI}
\tablenotetext{h}{$\sigma_{\mathrm{atmosphere}}$ is defined by Equation
\ref{sigAtmos} and depends on the projected baseline length and guide star
distance. We report the range of values for our 300-second blocks over one
night of observing.}
\tablenotetext{i}{\citet{Woillez10}}
\tablenotetext{j}{Derived using values from Table 3 of \citet{GRAVITY} and
includes contributions from the narrow angle baseline determination, the beam
combiner phase measurements, metrology, dispersion, and relativity.}
\ref{sigAtmos} and is a function of baseline. For the VLTI we report the mean
value for the six baselines. 
\tablecomments{Our error values are reported for 300 second blocks. Where
applicable they will average down. For example, in three hours,
$\sigma_{\mathrm{atmosphere}}$ for Keck Interferometer is $\sim7.2\mu$as. For the VLTI in six
hours, $\sigma_{\mathrm{atmosphere}}\sim1\mu$as \label{VisParams}}
\end{deluxetable}

\subsection{Observing Routine, UV-coverage, and PSFs}\label{ObsRoutine} 
We simulate observations for two instruments: ASTRA at the Keck Interferometer
and GRAVITY at the VLTI.  For the Keck Interferometer/ASTRA our adopted
observing routine assumes 10 visits to Sgr A* per night at even intervals
between the hour angles of -1.5 and 1.5. This is the maximum hour angle range
for which Sgr A* is above $\sim50^{\circ}$ Zenith Angle. For the VLTI, we
modeled 20 visits to Sgr A* per night between the hour angles of -3 and 3.
This reflects the higher transit of Sgr A* at the location of Cerro Paranal and
assumes a similar observational cadence is attainable at both the Keck
Interferometer and VLTI.  Individual observations are assumed to last 10
minutes with 5 minutes of on-source integration.  We model fringes in
5 spectral channels dispersed across the K-band (see Table \ref{specTable}).
Ten observations in five spectral channels per night provide 50 visibilities
from the Keck interferometer. 20 observations in five spectral channels over
6 baselines provide 600 visibilities per night at the VLTI.

\begin{figure}[h!] \begin{center}
\epsscale{0.8}
\plotone{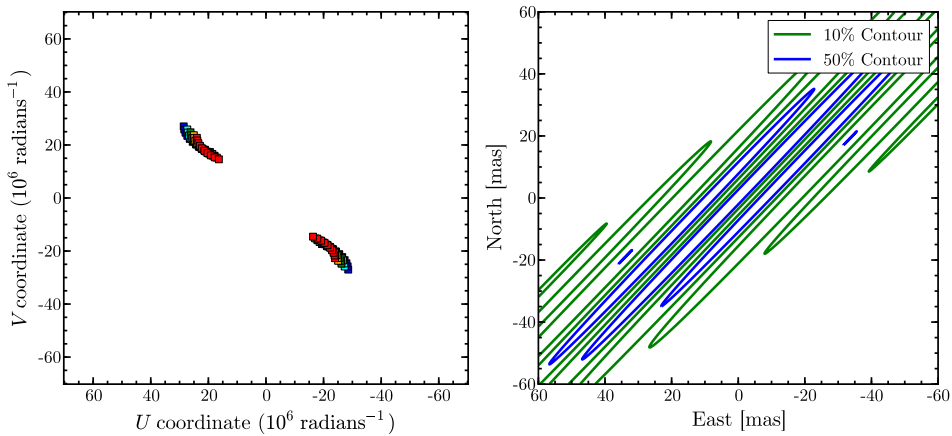}
\caption{The uv-coverage (left), and the 10\% and 50\% contours of the
resulting PSF (right) produced by our adopted observing routine at the Keck
Interferometer.\label{KeckBeam}}
\end{center}\end{figure}

The resulting uv-coverage and PSF for the Keck Interferometer is shown in Figure
\ref{KeckBeam}. The PSF is narrow, $\sim5$ mas at half maximum, along the direction of the
interferometer baseline, but extended in the perpendicular direction.  Because
of the distinct shape of the Keck Interferometer PSF, we will measure astrometry much more
precisely along the baseline than in the orthogonal direction. The extended
wings or sidelobes of the PSF will have the tendency to overlap when the
separation between sources has only a very small component along the baseline
direction. Such overlapping sidelobes bias astrometric measurements (see
below). However, within $\sim50$ mas of the Galactic Center, where stars are expected
to be orbiting with periods $\sim1$ year and $~1-2$ bright sources are expected
in the field, multi-epoch observations with the Keck Interferometer stand
a good chance of observing the sources with significant separation along the
baseline angle (see Section \ref{KeckResults} below).

The extended wings of the Keck Interferometer PSF shown in Figure \ref{KeckBeam} will result in
a restricted contrast limit. This is because flux from a bright source is
distributed throughout the field in the wings of the PSF. To be detectable in
the vicinity of a bright source, a faint source must be brighter than the noise
in the wings.

\begin{figure}[h!] \begin{center}
\epsscale{0.8}
\plotone{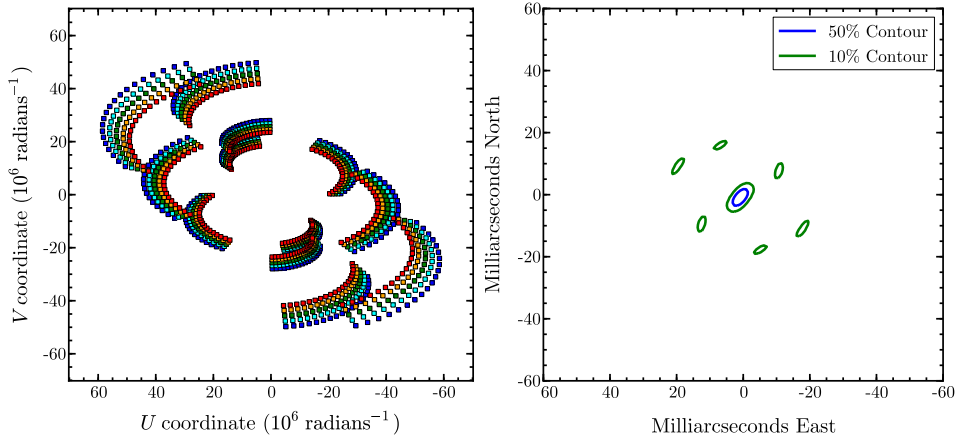}
\caption{
Left: The uv coverage provided by the VLTI in six hours of observing the
Galactic Center. Right: The 10\% and the 50\% contours
of the synthesized beam. \label{VLTBeam} }
\end{center}\end{figure}

Our adopted VLTI observing routine provides the uv-coverage shown in the left
panel of Figure \ref{VLTBeam}.  The increased uv-coverage provided
should have an easier time distinguishing sources in a crowded field.

\section{Fitting Stellar Positions to Visibility Data} 

There are four free parameters for each source in the field: position
($x$,$y$), flux, and spectral slope. For real data, the number of sources in
the field will not be known a priori.  In order to efficiently search the large
parameter space for a best fit to the synthesized data, we implement a hybrid
grid fit and Levenberg Marquardt (LM) minimization algorithm to minimize the
$\chi^2$ function of the parameters given the synthesized data.
 
\begin{figure}[h!] \begin{center}
\epsscale{0.5}
\plotone{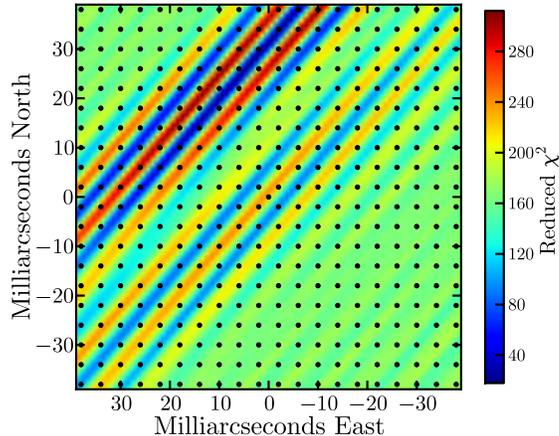}
\caption{ A slice through the reduced $\chi^2$ surface for a fit of
a two-source model to Keck Interferometer data. The undulations present
a hurdle to the naive application of a gradient fitting algorithm since there
is the potential to converge on a local minimum. Over plotted on the $\chi^2$
surface is the grid of seed positions provided to the gradient fitting
algorithm.  The grid spacing ensures that the global minimum is sampled.
\label{chiSurfKeck} }
\end{center} \end{figure}

\begin{figure}[h!] \begin{center}
\epsscale{0.8}
\plotone{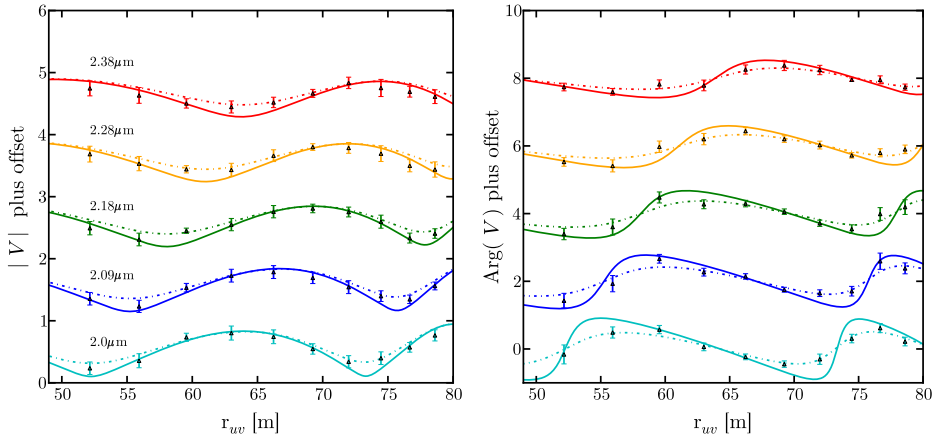}
\caption{The triangular points mark synthesized visibility data with measurement uncertainties for one
night of observing with ASTRA at the Keck Interferometer. We generate
50 complex visibility measurements over 3 hours of observation. Each
measurement is made with 300 seconds of on-source integration. The average
measurement uncertainty in both the amplitude and phase is $\sim15\%$. The
solid curve shows the visibility vs. baseline expected for our star field
ignoring instrumental effects. The dashed curve shows the visibility vs.
baseline for the star field with the source fluxes scaled by the Gaussian fiber
response function.\label{VisFig} } \end{center} \end{figure}

The interferometric visibility produced by a distribution of point sources is
expected to be undulating versus baseline (Figure \ref{VisFig}) and thus the
$\chi^2$ surface in the $x_k,y_k$ plane is expected to be similarly undulating
(see Figure \ref{chiSurfKeck} for a slice through the range of the $\chi^2$
function in the $x_k,y_k$ plane).  Since the Levenberg Marquardt algorithm is
a gradient fitting algorithm which always proceeds ``downhill" from an initial
starting point--- which must be supplied by the user--- the undulating nature
of the $\chi^2$ surface can present a problem because the LM algorithm will
converge on a local minimum if the initial parameters are not in the
neighborhood of the global minimum.  We have therefore adopted a grid-based
approach to ensure that starting values supplied to the LM fitter sample the
$\chi^2$ surface well enough to ensure at least one seed begins in the
neighborhood of the global minimum. Since local minima on the $\chi^2$ surface
mirror local maxima in the PSF, local minima will be separated by about the
interferometer fringe spacing.  Thus, we used a grid spacing of
$\frac{\lambda_{\mathrm{min}}}{B_{\mathrm{max}}}$, where
$\lambda_{\mathrm{min}}$ is the shortest wavelength observed and
$B_{\mathrm{max}}$ is the longest projected baseline used. Figure
\ref{chiSurfKeck} shows our grid of seed parameters.

To begin, we start with a model consisting of two point sources. For the fit,
we seed an LM fitter with a grid of starting locations (shown in Figure
\ref{chiSurfKeck}) and a guess of the flux and spectral slope of each source.
We then use the best-fit values of this fit together with an additional grid of
starting locations to seed a 3-source model fit. This process can then be
repeated to fit for any number of sources. We do not know how many sources will
be present in real data, so we are guided by the significance of the fits and
the deduced flux of the fitted sources. Highly significant fits and sources
with large fitted fluxes are taken to be real, and less significant fits are
disregarded.

\section{Results\label{ResultsSec}}
\subsection{Star Fields Observed with ASTRA at the Keck Interferometer\label{KeckResults}} 
\begin{figure}[h!] \begin{center}
\epsscale{0.5}
\plotone{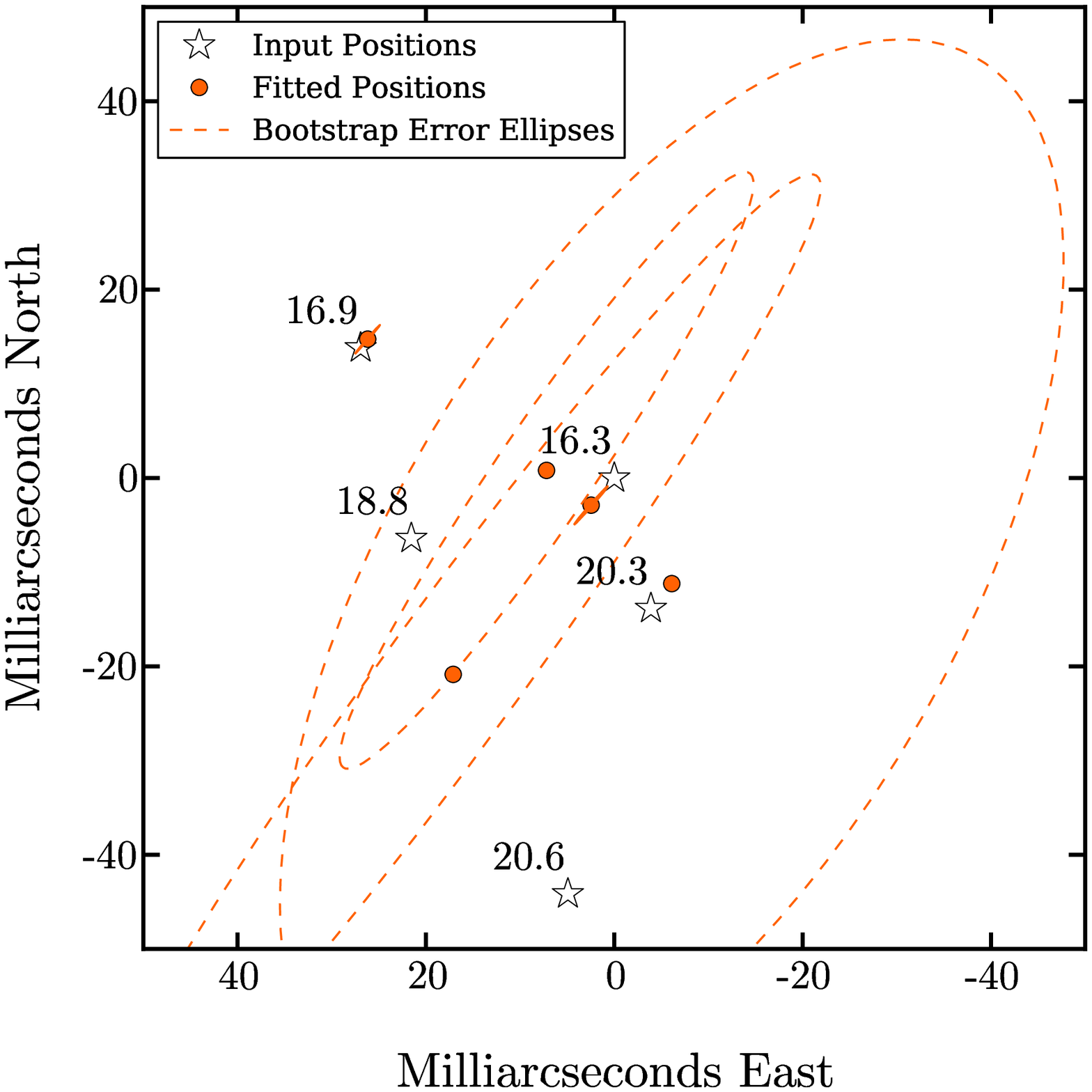}
\caption{
One epoch of the orbiting star field Field1 (white star symbols). Our recovered
positions and astrometric error ellipses are also plotted (orange symbols).
Only the positions of Sgr A* and Star 1 are accurately recovered.  The
positions of the fainter sources are not recovered. We have plotted the error
ellipses of these sources using a dashed line. Note that these dashed curves
are error ellipses, and not orbits. \label{1epoch1mag}}
\end{center}\end{figure}

For our simulated Keck Interferometer observations of the orbiting star field
Field1,  we adopt a two-year observing routine that includes two three-night
observing runs per year, one in the spring and one in the late summer. In
Figure \ref{1epoch1mag} we show our fitted source positions as well as the
input positions for one of the nights.  The positions of Sgr A* at $m_K=16.3$
and of Star 1 at $m_K=16.9$ are recovered (small error ellipses), but the
positions of the fainter stars are not recovered (very large error ellipses)
and are hereafter disregarded. 

\begin{figure}[h!] \begin{center}
\epsscale{0.8}
\plottwo{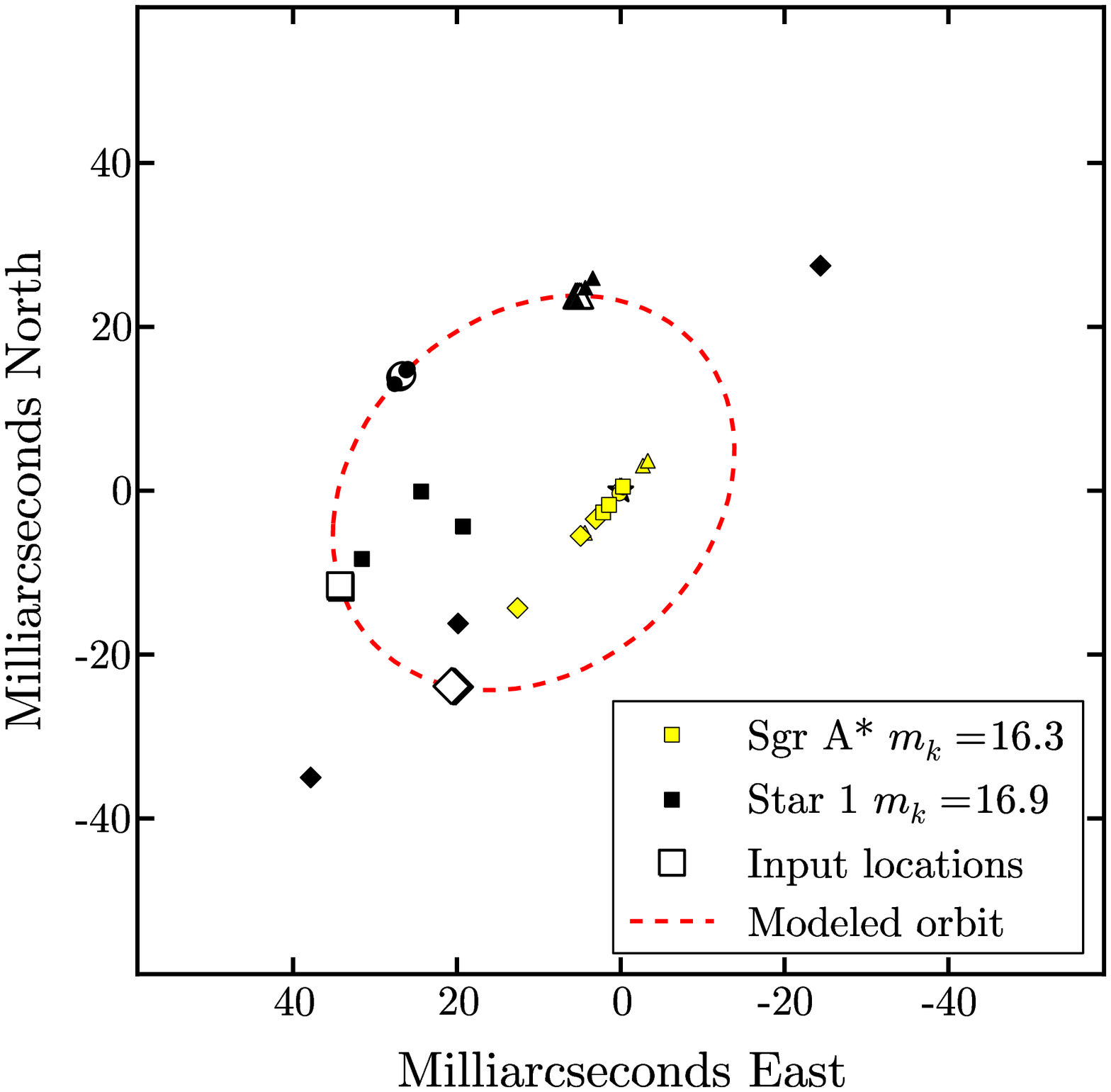}{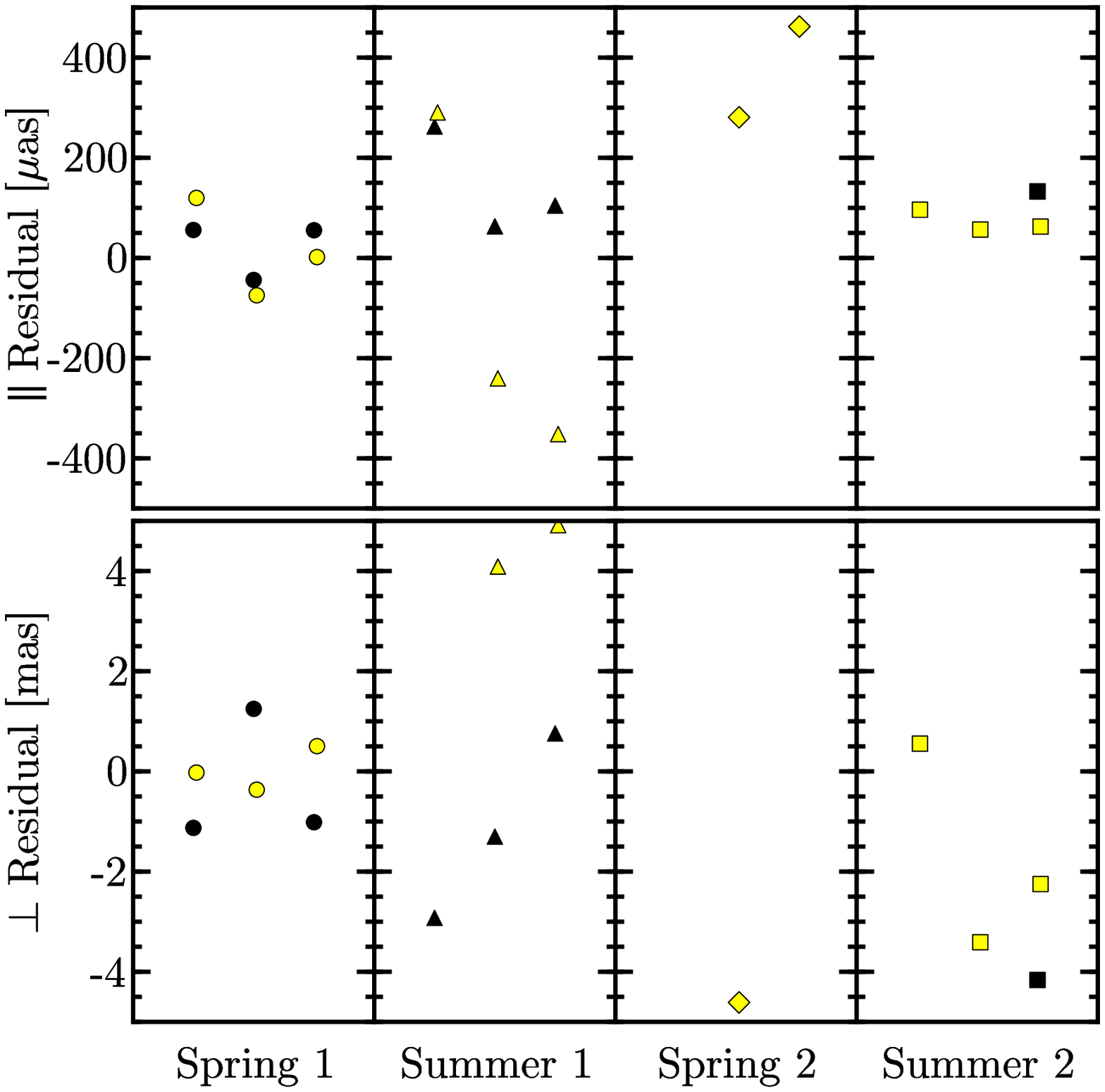}
\caption{
Left: We show the recovered positions of Sgr A* (yellow symbols) and Star
1 (black symbols) for four three-night runs of simulated Keck Interferometer
data. The fits from each run are plotted with a unique symbol shape. The input orbit of
Star 1 is also plotted (red dashed line) together with the expected location of
Star 1 for each epoch (larger white symbols). Right: The astrometric residuals along
the average baseline direction (top row) and the perpendicular direction
(bottom row). Note the change from microarcseconds in the top row, to
milliarcseconds in the bottom row.\label{Field1Orbit}} \end{center}\end{figure}

\begin{figure}[h!] \begin{center}
\epsscale{0.8}
\plotone{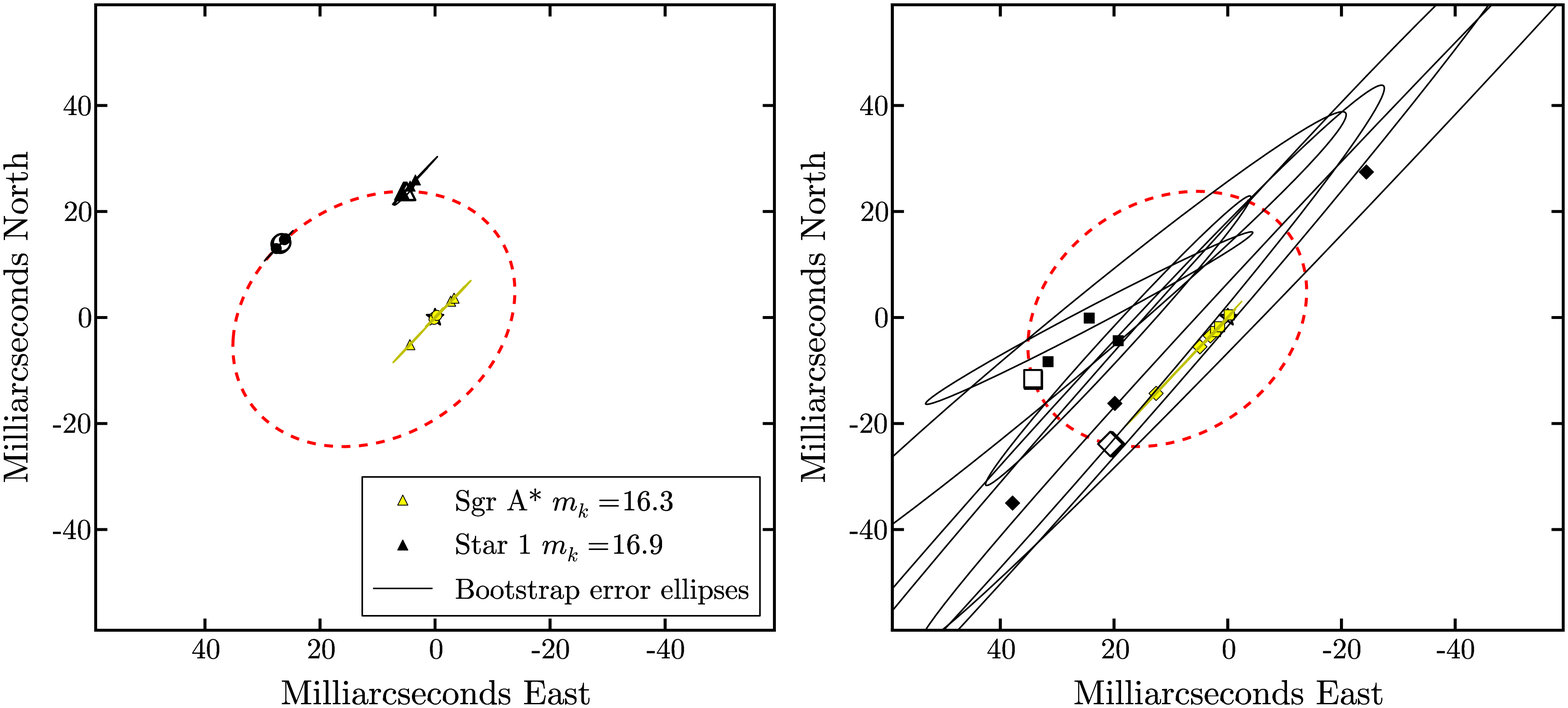}
\caption{We have split the left panel of Figure \ref{Field1Orbit} into two
parts showing the recovered positions from the first year of data in the left
panel, and the recovered positions for the second year of data in the right
panel. Astrometric uncertainties for each fitted point are indicated with
a solid line.  Due to the large sidelobes in the Keck Interferometer PSF
(Figure \ref{KeckBeam}) our recovered positions are most precise when Star
1 and Sgr A* are well separated along the direction of the interferometer
baseline.\label{Field1Select}} \end{center}\end{figure}

Figure \ref{Field1Orbit} shows the fitted position for Sgr A* and Star 1 for
all 12 nights of observing Field1. The fainter stars have not been plotted
because their positions are not recovered by the fitter.  The astrometric
residuals for Sgr A* and Star 1 are shown in the right panel of Figure
\ref{Field1Orbit}.  The astrometric precision in our fits can be estimated in
three ways. First, we are able to characterize the quality of fits by computing the
true residuals (right panel of Figure \ref{Field1Orbit}).  Second, by assuming
no significant orbital motion over the three-day time period, the dispersion in
fitted positions over consecutive nights provides a rough measure of the
astrometric precision for each observing run. Lastly, in Figure
\ref{Field1Select} we show calculated error ellipses generated using a standard
bootstrapping algorithm. 

Bootstrapping randomly re-samples the data, with replacement, several times and
re-runs the fitter on each new sample. The process of drawing and replacing
ensures that for most resampled data sets some of the data is redundant and
some of the original data is missing. When our source fitter is run on the
re-sampled data sets, a range of fitted parameter values are returned. The
spread in the returned parameter values defines the shape of the uncertainty
ellipses.

In Figure \ref{Field1Select} we split the recovered positions shown in the left
panel of Figure \ref{Field1Orbit} into two plots, the left showing the results
from the first year and the right plot showing the results from the second year
of observation. We also show the astrometric error ellipses derived via our
bootstrapping routine. As expected based on the shape of the Keck Interferometer PSF (Figure
\ref{KeckBeam}), our ability to accurately recover the position of the star
depends on the position angle between the star and Sgr A*. During the first
year (left panel of Figure \ref{Field1Select}), Star 1 is well separated in the
direction of the baseline from Sgr A* and the astrometric residuals are
$\sim100\mu$as along the baseline direction and $\sim4$mas in the perpendicular
direction. In the second year (right panel) when Star 1 and Sgr A* are not well
separated in the direction of the interferometer baseline, our astrometry is
poor as indicated by the larger spread in fitted positions over the three-night
runs, the larger residuals, and the significantly larger error ellipses derived
for the epochs shown in \ref{Field1Select}.  This is due, as discussed above,
to overlapping sidelobes.  The distinct shape of the bootstrap error
ellipses which are much narrower in the direction of the interferometer
baseline than in the direction perpendicular, reflects the shape of the Keck
Interferometer PSF which has similar features.

\begin{figure}[h!] \begin{center}
\epsscale{0.5}
\plotone{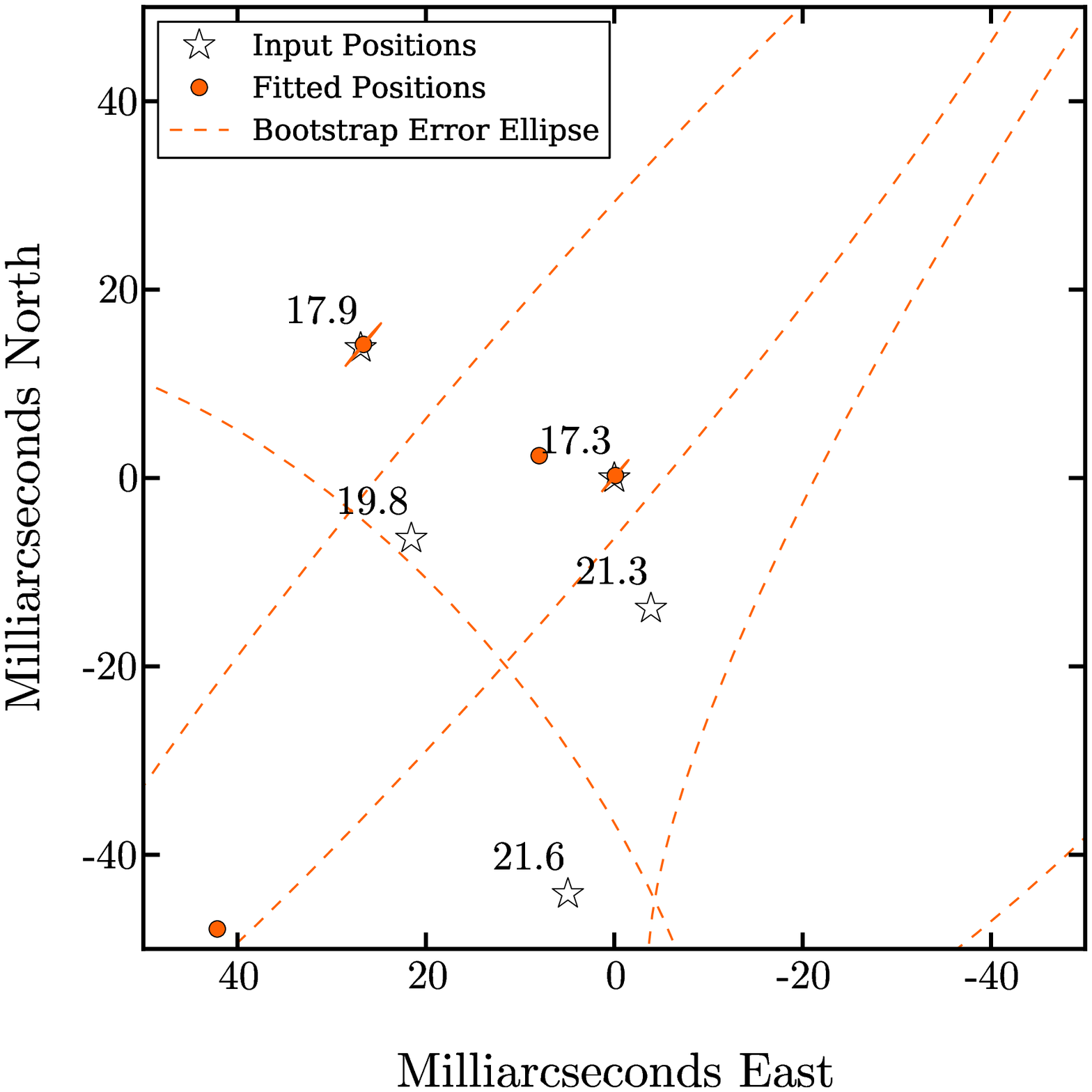}
\caption{
This plot is the same as Figure \ref{1epoch1mag} but for Field2. We are unable
to recover the positions of the fainter stars. The astrometric error ellipses
for the unrecovered sources are plotted as dashed lines; note that these are
not orbits.  \label{1epoch2mag} }
\end{center}\end{figure}

We adopt the same observing program as for Field1, namely four three-night runs
over two years for our simulated observations of Field2.  Our fit to one
night's data is shown in Figure \ref{1epoch2mag}. We
recover the position of Sgr A* and Star 1 but we are unable to recover the
positions of the fainter stars.

\begin{figure}[h!] \begin{center}
\epsscale{0.8}
\plottwo{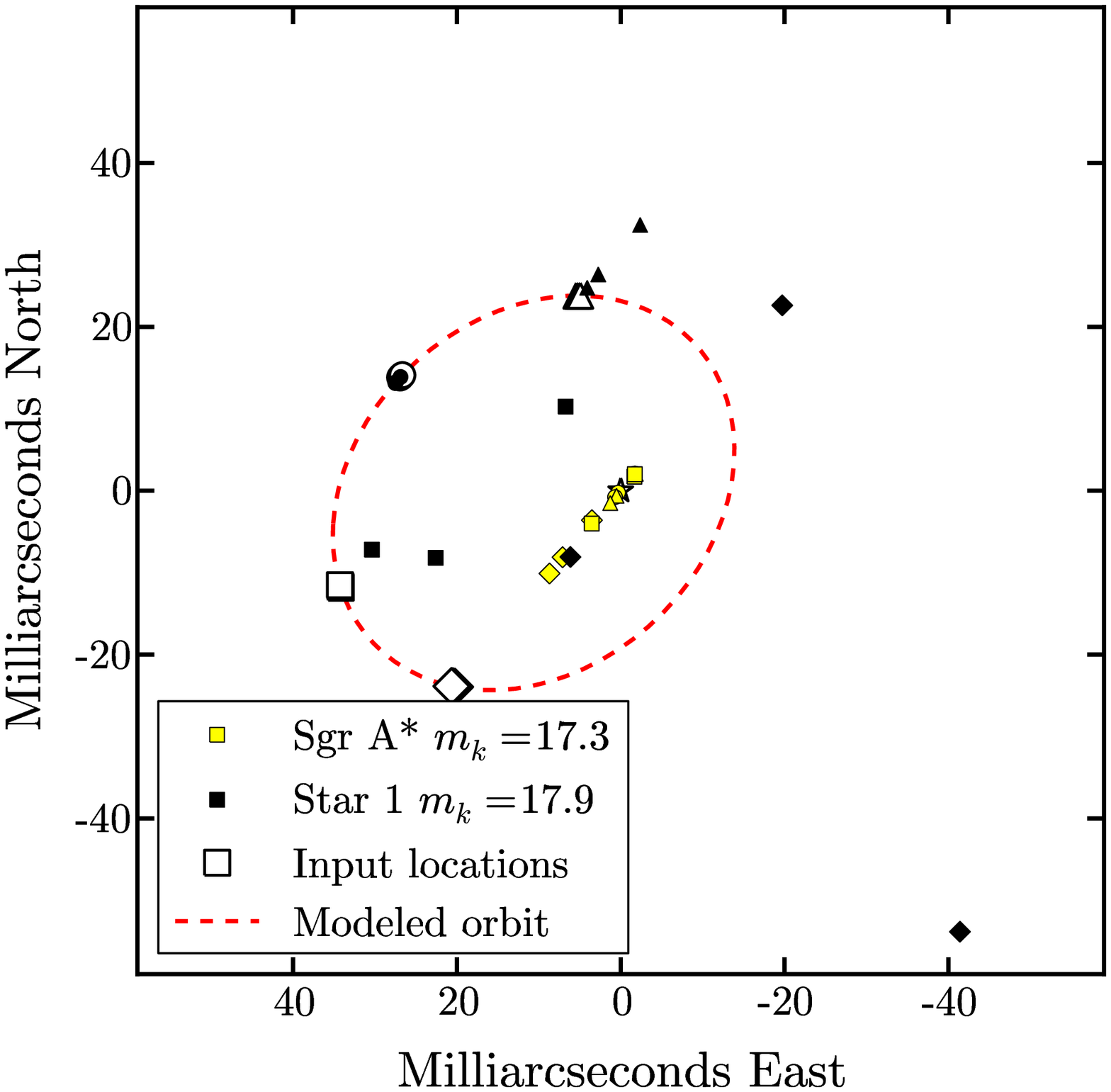}{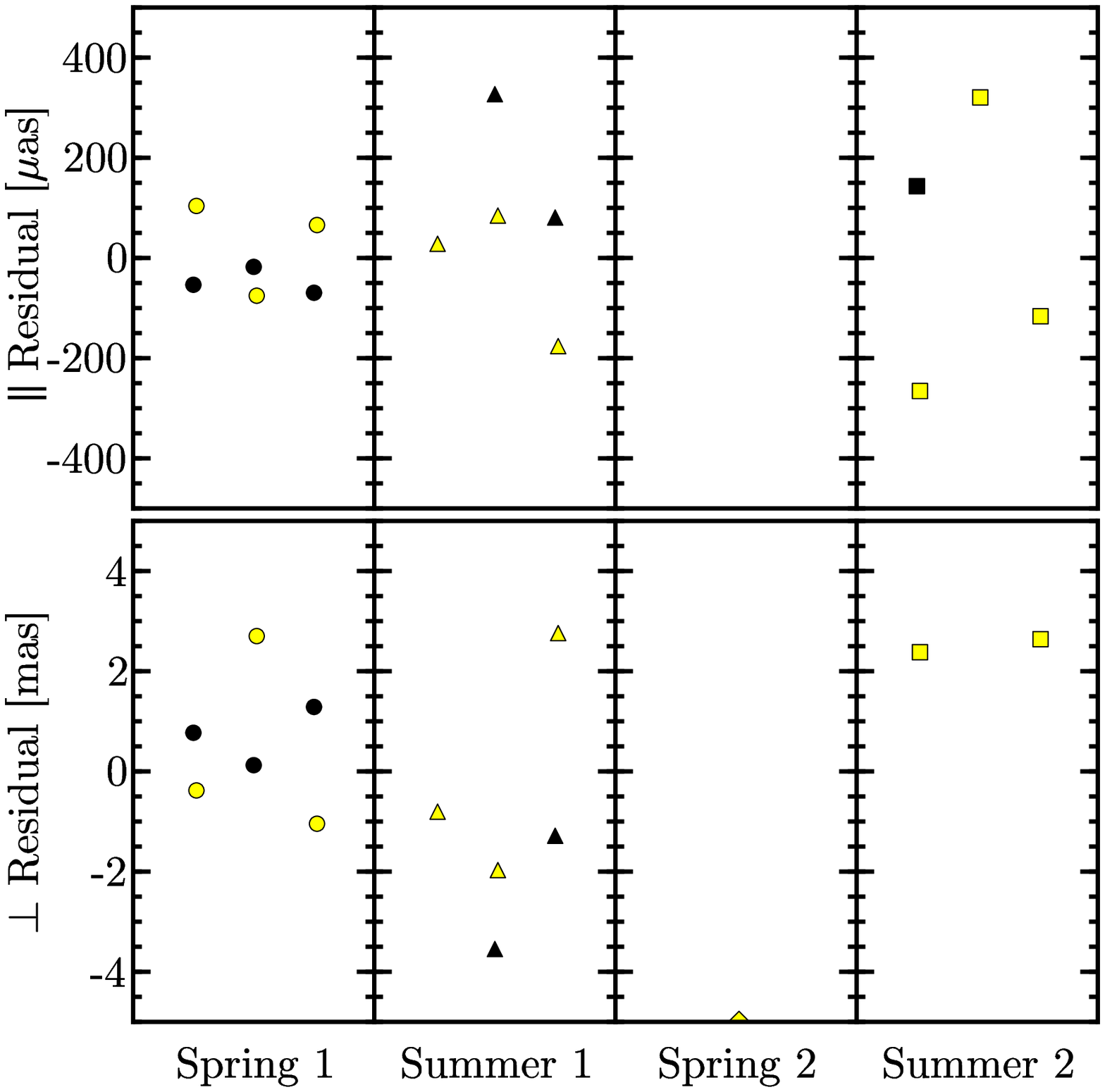}
\caption{Left: Our recovered astrometric positions for each night of observing
Field2. Black symbols represent Star 1 recovered positions, and yellow symbols
refer to Sgr A* recovered positions.  Symbol shapes are unique to each of the
four three-day observing runs. Right: The astrometric residuals for each night
are shown. Along the baseline direction the residuals are plotted in
microarcseconds, while in the perpendicular direction they are plotted in
milliarcseconds.  \label{Field2Orbit}} \end{center}\end{figure}

\begin{figure}[h!] \begin{center}
\epsscale{0.8}
\plotone{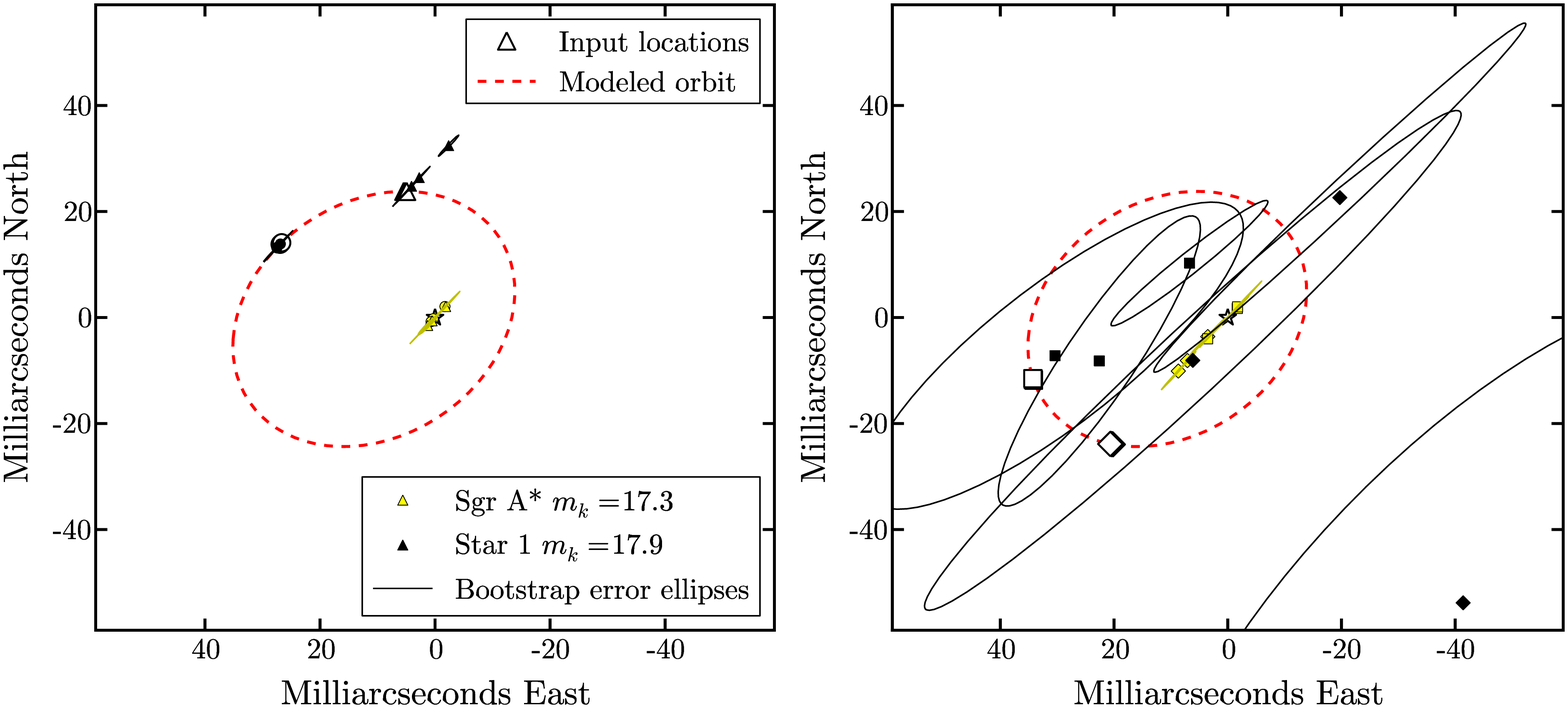}
\caption{In this plot we split the left panel of Figure \ref{Field2Orbit}
showing the data from the first year in the left plot, and the data from the
second year in the right plot. Black symbols still refer to Star 1 recovered
positions, and yellow symbols refer to Sgr A* recovered positions. The symbol
shapes designate the observing run in the same way as in Figure
\ref{Field2Orbit}. Solid curves indicate the astrometric uncertainty. As
discussed, our astrometry is worse in the second year due to overlapping
sidelobes.\label{Field2Select}} \end{center}\end{figure}

Figure \ref{Field2Orbit} shows our fitted positions for Sgr A* and Star 1 for
each of the 12 nights of observing Field2.  As is the case for Field1, our
ability to recover the positions of Sgr A* and Star 1 is hindered when the
sidelobes of each source overlap in the Spring and Summer of the second year.
In Figure \ref{Field2Select} we split the left panel of Figure
\ref{Field2Orbit} into two panels, showing the data from each year separately.
For the observations during the first year our astrometric residuals on Sgr A*
($m_K=17.3$) and Star 1 ($m_K=17.9$) are $\sim200~\mu$as along the baseline
direction and $\sim4~\mathrm{mas}$ in the perpendicular direction (first two
columns in the right panel of Figure \ref{Field2Orbit}). 

To generate a complete picture of why we are unable to recover the fainter
sources, we investigate: 1) the signal-to-noise ratio of each source; and 2)
source confusion, which incorporates source density and source contrast.

In Table \ref{SNField1} we show an upper limit to the signal-to-noise ratio for
each source in Field1. To compute the values in Table \ref{SNField1} we made
some simplifying assumptions to help elucidate some of the issues without the
application of a more opaque formal treatment.  Namely, we assume that there is no flux
attenuation from the optical fiber and that each source is the only source
present in the field.  These two assumptions imply that our signal-to-noise
ratios are strict upper limits.  For example, a source with a reported upper
limit to the signal-to-noise ratio of 10 or greater may provide no detectable
signal in the presence of photon noise from brighter nearby sources or if the
signal of an-off-axis source is attenuated by the optical fiber response.
Additionally, including only one source in the field assumes a maximum
visibility amplitude. With multiple sources in the field the fringe signal is
diminished and the signal-to-noise ratio of the fringes of the more complex
star field will be similarly reduced. In the limit of a crowded and complex
star field, even large values in Table \ref{SNField1} do not necessarily imply
a high signal-to-noise in simulated data. However, small values do ensure
non-detections.

We list signal-to-noise values for each source with and without injection
fluctuations. The large change in the upper limit to the signal-to-noise ratio
when injection fluctuations are included indicate that they introduce a large
source of noise. Since even the upper limits indicate a marginal
signal-to-noise ratio for Stars 3 and 4, these sources are likely undetectable
in the presence of the brighter sources included in our actual simulated data. 
\begin{deluxetable}{ccccccc}
\tabletypesize{\scriptsize}
\tablewidth{0pt}
\tablecaption{Keck simulated data signal to noise for each spectral channel for Field1}
\tablehead{
    \colhead{Spectral Channel} &
    \colhead{Sgr A* ($m_K=16.3$) } & 
    \colhead{Star 1 ($m_K=16.9$) } & 
    \colhead{Star 2 ($m_K=18.8$) } & 
    \colhead{Star 3 ($m_K=20.3$) } & 
    \colhead{Star 4 ($m_K=20.6$) }  
    }
\startdata
\sidehead{Without injection fluctuations}
2.00 microns& 246 & 164 & 30 & 7 & 6 \\ 
2.09 microns& 242 & 146 & 26 & 6 & 5 \\ 
2.18 microns& 220 & 120 & 21 & 5 & 4 \\ 
2.28 microns& 200 &  94 & 17 & 4 & 3 \\ 
2.38 microns& 230 & 172 & 32 & 8 & 6 \\ 
\sidehead{With injection fluctuations}
2.00 microns&  81 &  77 & 29 & 7 & 6 \\ 
2.09 microns&  81 &  75 & 25 & 6 & 4 \\ 
2.18 microns&  80 &  70 & 21 & 5 & 4 \\ 
2.28 microns&  78 &  64 & 16 & 4 & 3 \\ 
2.38 microns&  80 &  78 & 30 & 8 & 6 \\ 
\enddata
\tablecomments{
This table shows the simulated signal-to-noise values for the sources in
Field1. As discussed in the text these values are upper limits and are included
here to give a rough feeling of the sensitivity limits. Injection fluctuations
dominate the noise for the brighter sources while background and readnoise are
significant for the fainter sources. As a convenience, we include the simulated
signal-to-noise ratio constructed without any injection fluctuations for
reference.  \label{SNField1} } \end{deluxetable}

Table \ref{SNField2}, like Table \ref{SNField1} for Field1, illustrates our
upper limits to the signal-to-noise ratio simulated for each source
in Field2.  Even the upper limits on the signal-to-noise ratio indicate that no
real signal is detected for Stars 3 and 4.
\begin{deluxetable}{ccccccc}
\tabletypesize{\scriptsize}
\tablewidth{0pt}
\tablecaption{Keck simulated data signal to noise for each spectral channel for Field2}
\tablehead{
    \colhead{Spectral Channel} &
    \colhead{Sgr A* (k=17.3) } & 
    \colhead{Star 1 (k=17.9) } & 
    \colhead{Star 2 (k=19.8) } & 
    \colhead{Star 3 (k=21.3) } & 
    \colhead{Star 4 (k=21.6) }  
    }
\startdata
 \sidehead{Without injection fluctuations}
 2.00 microns& 108 &    70 &    12 &     3 &     2 \\ 
 2.09 microns& 103 &    61 &    10 &     3 &     2 \\ 
 2.18 microns&  91 &    48 &     9 &     2 &     2 \\ 
 2.28 microns&  82 &    39 &     7 &     2 &     1 \\ 
 2.38 microns& 102 &    73 &    13 &     3 &     2 \\ 
\sidehead{With injection fluctuations}
 2.00 microns&  68 &    55 &    12 &     3 &     2 \\ 
 2.09 microns&  67 &    49 &    11 &     3 &     2 \\ 
 2.18 microns&  64 &    42 &     9 &     2 &     2 \\ 
 2.28 microns&  60 &    36 &     7 &     2 &     1 \\ 
 2.38 microns&  67 &    56 &    13 &     3 &     3 \\ 
\enddata
\tablecomments{
Simulated upper limits to the signal-to-noise ratio for each source in Field2.
\label{SNField2} }
\end{deluxetable}

\begin{figure}[h!] \begin{center}
\epsscale{0.5}
\plotone{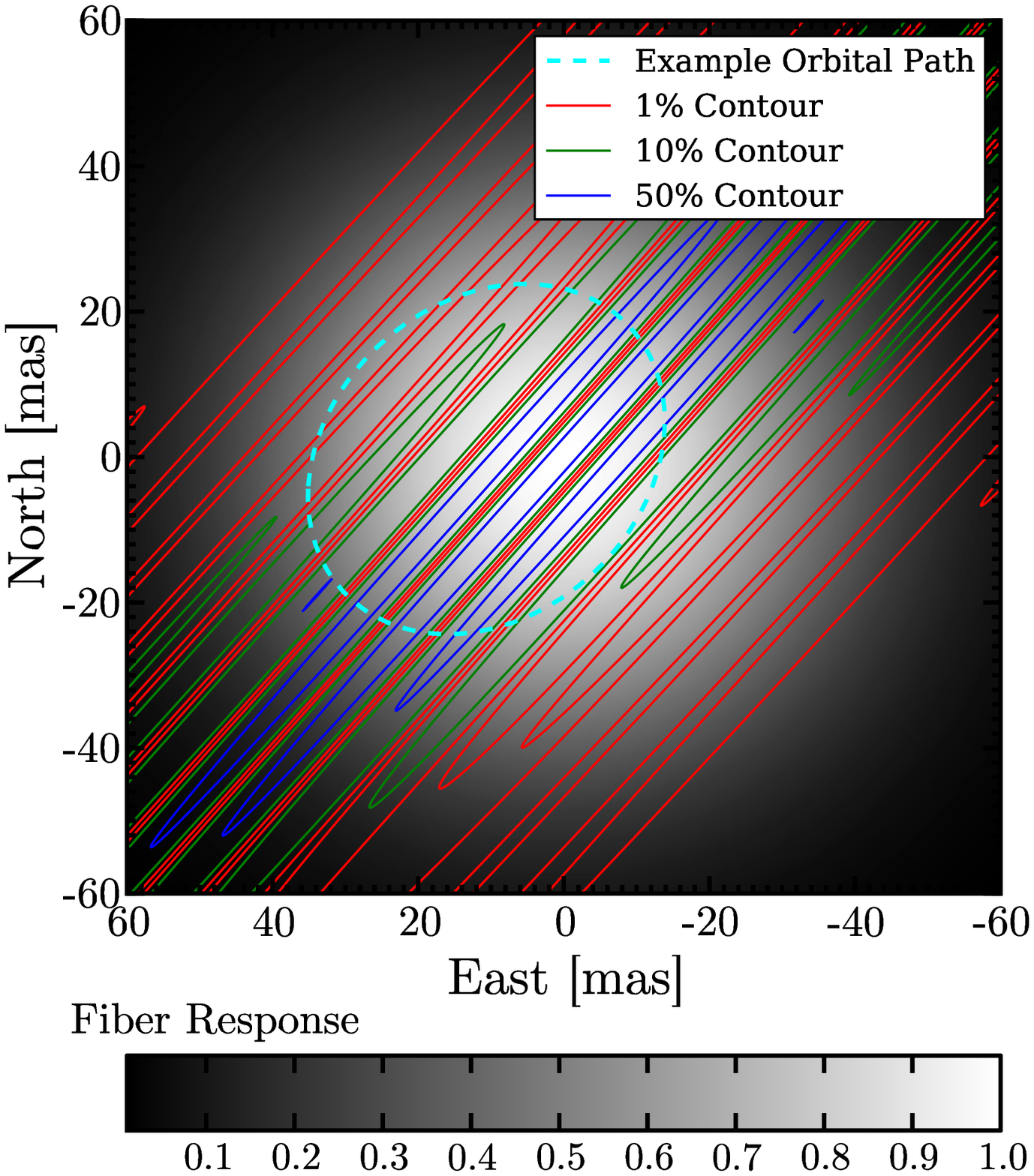}
\caption{In order to illustrate some of the difficulties in detecting a stellar
source in orbit about Sgr A* we overplot one of our adopted stellar
orbits (cyan dashed curve) on top of the Keck Interferometer PSF (solid red,
green, and blue contours at 1\%, 10\%, and 50\% power respectively). We
also show the optical fiber response in grayscale.
\label{KeckBeamOrbit}}
\end{center}\end{figure}

As we discussed in Section \ref{ObsRoutine} the sidelobes of the Keck
Interferometer PSF will impose a confusion limit in the Keck Interferometer
data both because the lobes will set a contrast limit and because they will
tend to overlap when sources are not well separated. In Figure
\ref{KeckBeamOrbit} we show the 1\% (red), 10\% (green), and 50\% (blue)
contours of the Keck Interferometer PSF. We also plot the fiber response
function and the orbital path of one of our stars. This plot shows that
detecting a faint source will be easiest when the star enters a region where
the sidelobe flux from Sgr A* is lowest.  However, there is the competing
factor of the fiber response function which tends to attenuate the flux from
sources which are located far from the center of the field. Thus while there
are some regions in the field beyond the 1\% contours of the PSF, detecting
a source there is made difficult by the low transmission of the fiber.

\begin{figure}[h!] \begin{center}
\epsscale{0.8}
\plotone{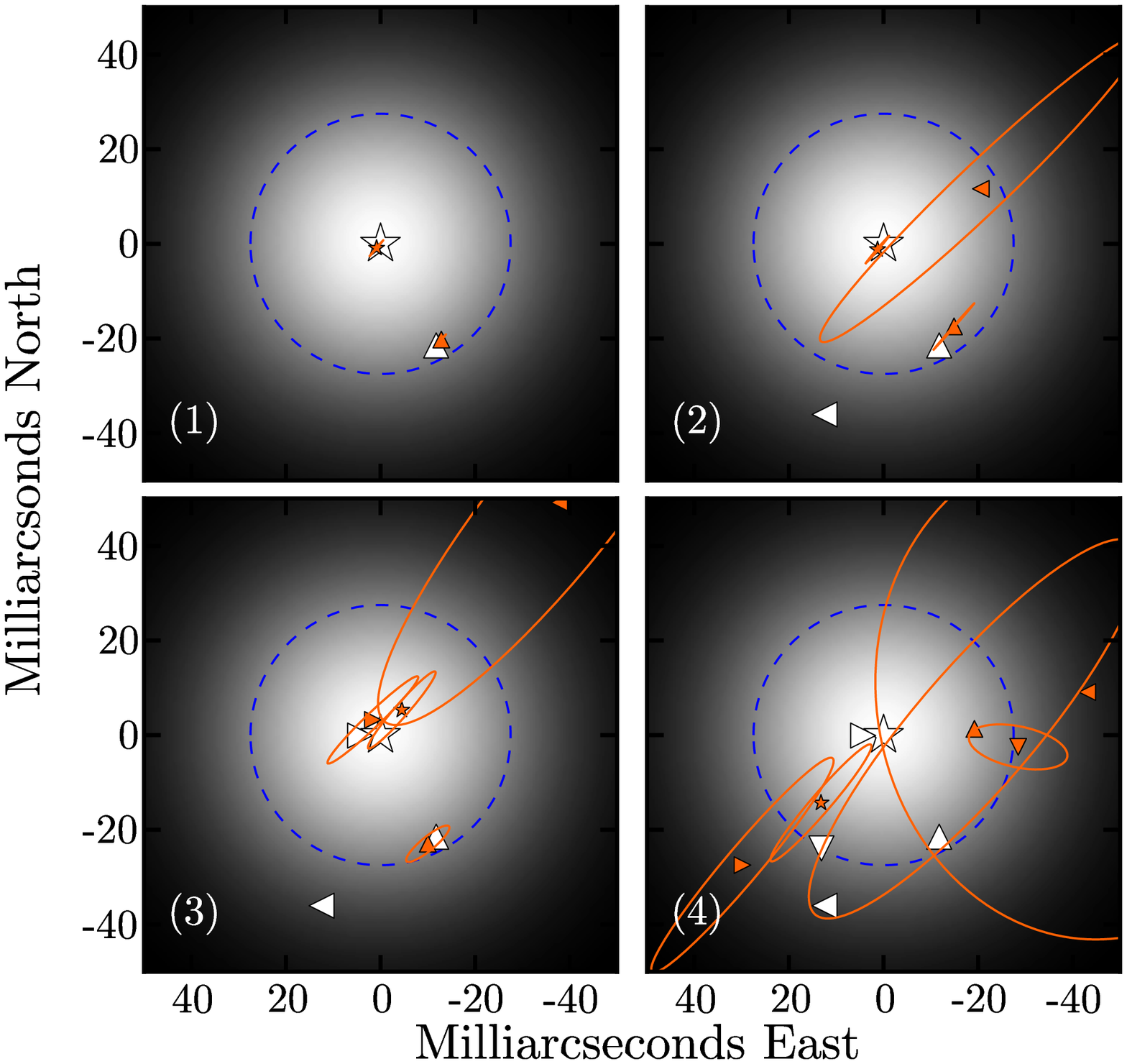}
\caption{ 
In these panels our recovered positions (orange symbols) and error ellipses are
plotted showing our performance recovering source positions from Keck Interferometer data when
more and more sources are present. The input star field for each panel includes
Sgr A* at the origin with $m_K=16.3$ (star symbol). In addition to Sgr A* each
star field also includes from 1 to 4 $m_K=17$ stars (isosceles triangles with
vertices pointing up, right, left, and down corresponding to the first, second,
third and fourth added star respectively). Also plotted is the Gaussian fiber
response function (grayscale) and the 50\% contour of this function (blue
dashed line). Source fluxes are attenuated by this function before detection.
\label{confusionFig}} \end{center}\end{figure}

An independent limit distinct from contrast but prominent in confusion noise
is source crowding.  In an attempt to isolate the effects of crowding and
provide evidence of whether crowding is limiting all the previous fits,
we also simulate data for four star fields, each with a different number of
$m_K=17$ stars. These simulations also provide some insight into the potential
performance of the Keck Interferometer if the KLF at the Galactic Center is
significantly flatter than Field1 or Field2. The four panels of Figure
\ref{confusionFig} show our fits to these star fields.  The star fields were
constructed as follows: First, one source with $m_K=16.3$ is placed at the origin
and another with $m_K=17.0$ is placed randomly within a 100 x 100
milliarcsecond field (panel one). To this star field, an additional $m_K=17.0$
source is placed randomly in the field (panel two), and so on until a total of
four $m_K=17.0$ sources are present in addition to the central $m_K=16.3$
magnitude source (panels 3 and
4).  

In panel one of Figure \ref{confusionFig}, our fits to the simulated visibility
data recover the input positions of both sources. Our confidence in these fits
is implied by the small error ellipses generated by our bootstrapping routine.
In panel two, a second $m_K=17$ magnitude source has been added to the star
field beyond the half-maximum radius of the optical fiber attenuation function.
Note that due to the fiber function, the flux of the unrecovered star is
attenuated by $\sim1$ magnitude. We are still able to recover the positions of
the first two sources but we cannot recover the position of this third source
due mostly to the increased contrast caused by the fiber-function-attenuated
flux. In the third panel, a third $m_K=17$ magnitude source is added to the
field, this time very near the origin. Our fits to this field do recover the
positions of the three sources within the half-maximum radius of the fiber
function with some confidence; the star placed outside this radius is still not
recovered. 

The accuracy and the precision of the recovered source positions in panel 3 are
somewhat degraded compared to the recovered positions in panels 1 and 2.  This
degradation in precision is due to the increased crowding of the field with
relatively bright sources and the effects of overlapping sidelobes. In the
fourth panel of Figure \ref{confusionFig}, with 5 bright sources in the field,
the effects of overlapping sidelobes are so severe that we cannot recover the
position of any source.

\subsection{Star Fields Observed with GRAVITY} 
\begin{figure}[h!] \begin{center}
\epsscale{0.8}
\plottwo{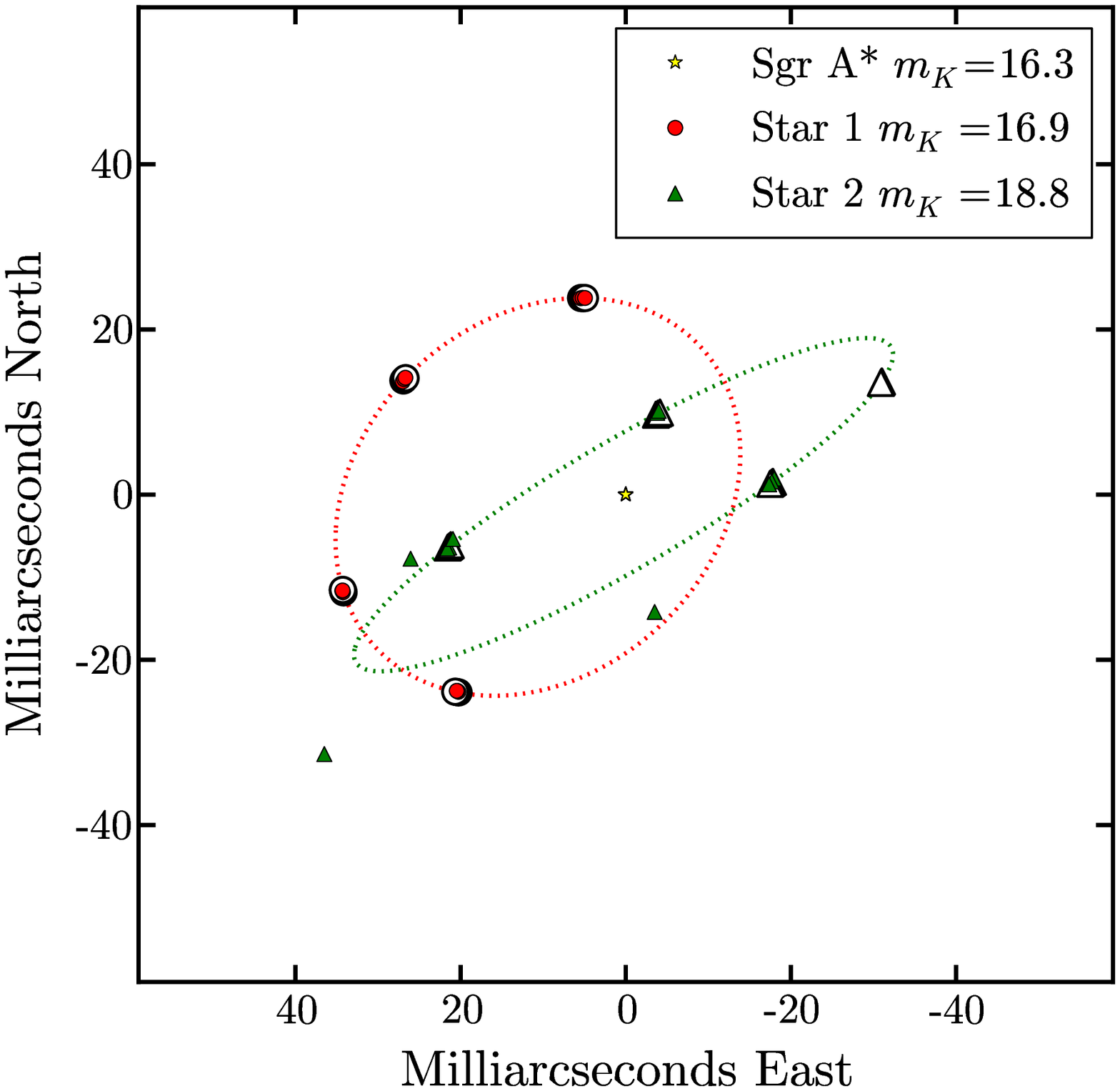}{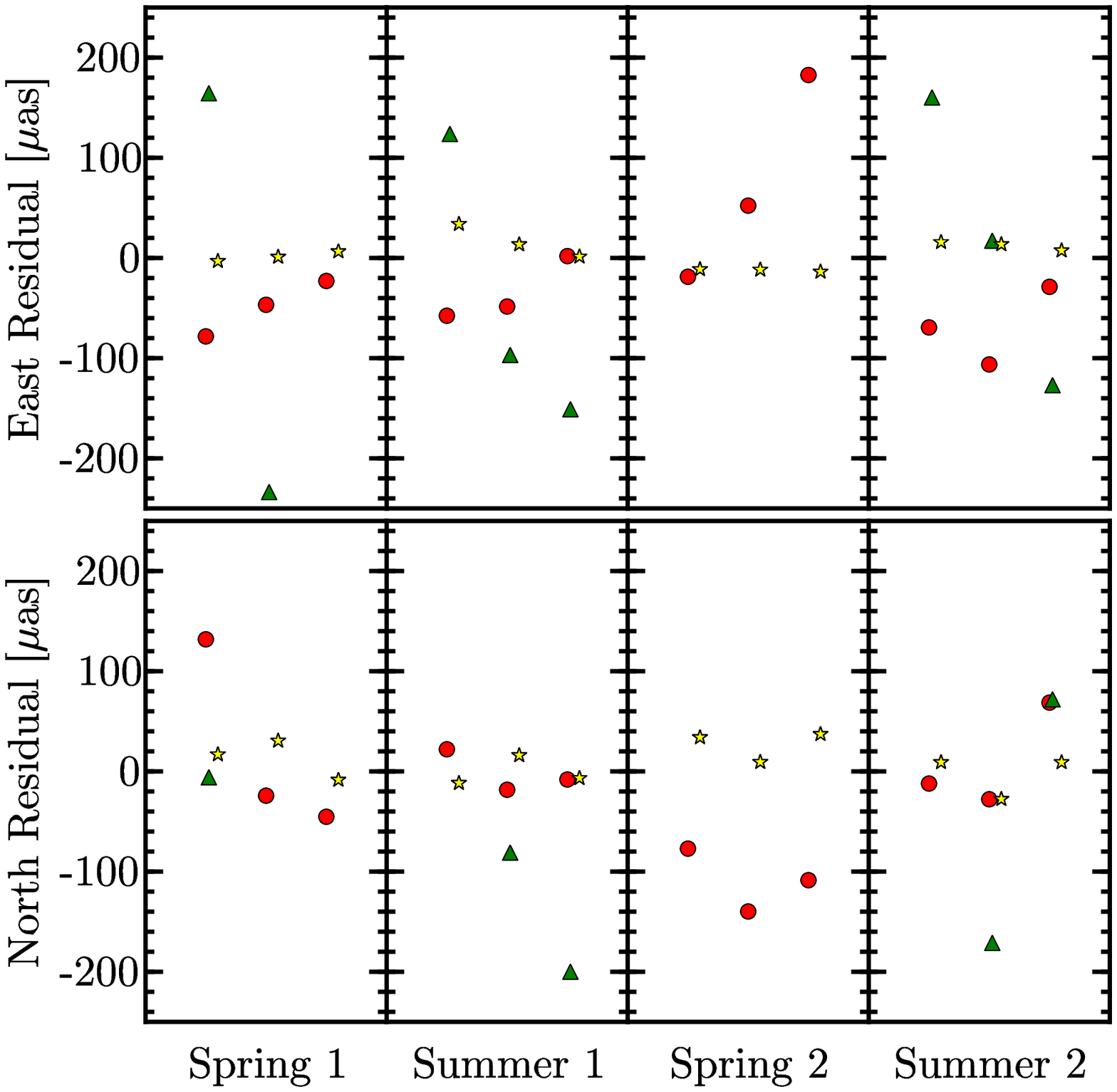}
\caption{
This plot is similar to Figure \ref{Field1Orbit} but shows the results for our
GRAVITY simulations.  Left: The recovered positions of Sgr A* (yellow star
symbols), Star 1 (red circle symbols), and Star 2 (green triangle symbols) for
12 nights of simulated GRAVITY data.  The input orbits of Star 1 and Star 2 are
also plotted (red and green dashed lines respectively).  White symbols show the
input location for each source. Note that in our GRAVITY plots, symbol shape is
used to identify sources, not observing epoch. Right: The astrometric residuals
in the North and East directions are plotted for both Sgr~A*, Star~1, and
Star~2.  \label{VLTField1Orbit} } \end{center}\end{figure}

\begin{figure}[h!] \begin{center}
\epsscale{0.5}
\plotone{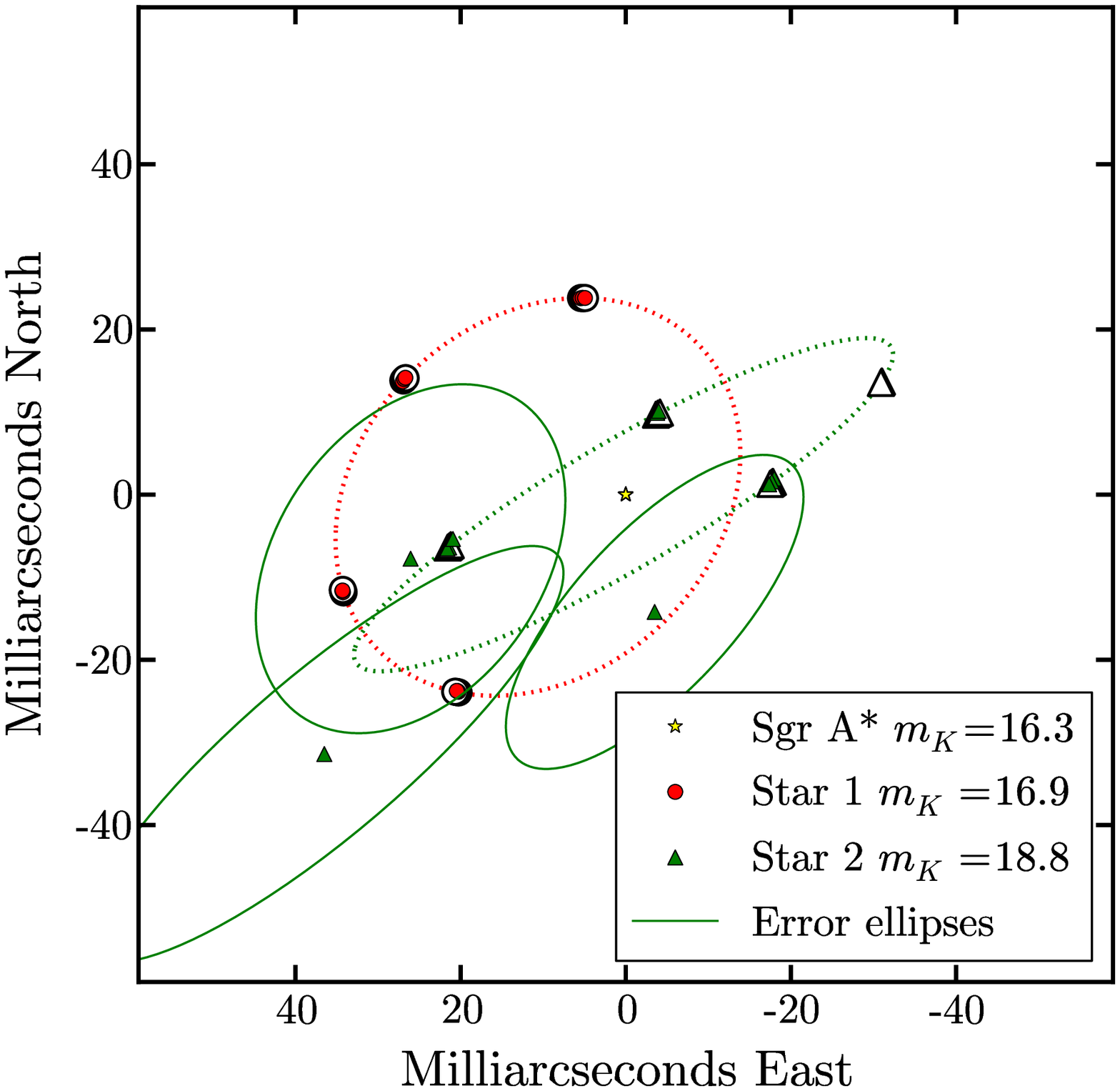}
\caption{The left panel of Figure \ref{VLTField1Orbit} but here we include the
bootstrap error ellipses. Where ellipses cannot be seen they are smaller than
the plotting symbols. White symbols indicate input locations.\label{VLTField1Ellipses}}
\end{center}\end{figure}

Figure \ref{VLTField1Orbit} shows the recovered positions for Sgr A*, Star 1,
and Star 2 for four three-night observing runs following the same schedule that
was used for the Keck simulations. The astrometric residuals (shown in the
right panel of Figure \ref{VLTField1Orbit}) are $\sim10~\mu$as,
$\sim100~\mu$as, and $\sim200~\mu$as for Sgr A*, Star 1, and Star
2 respectively. In Figure \ref{VLTField1Ellipses} we also show the bootstrap
error ellipses associated with our fitted positions; where none are seen they
are smaller than the plotting symbols.  Stars 3 and 4 are not plotted because
their positions are not well recovered.

For one run, when Star 2 is farthest from the center of the field, we are
unable to recover its position on any of the three nights.  The optical
fiber transmission function attenuates the flux most during this run. We
investigate some of the limiting factors to recovering source positions in
GRAVITY data below.

\begin{figure}[h!] \begin{center}
\epsscale{0.8}
\plottwo{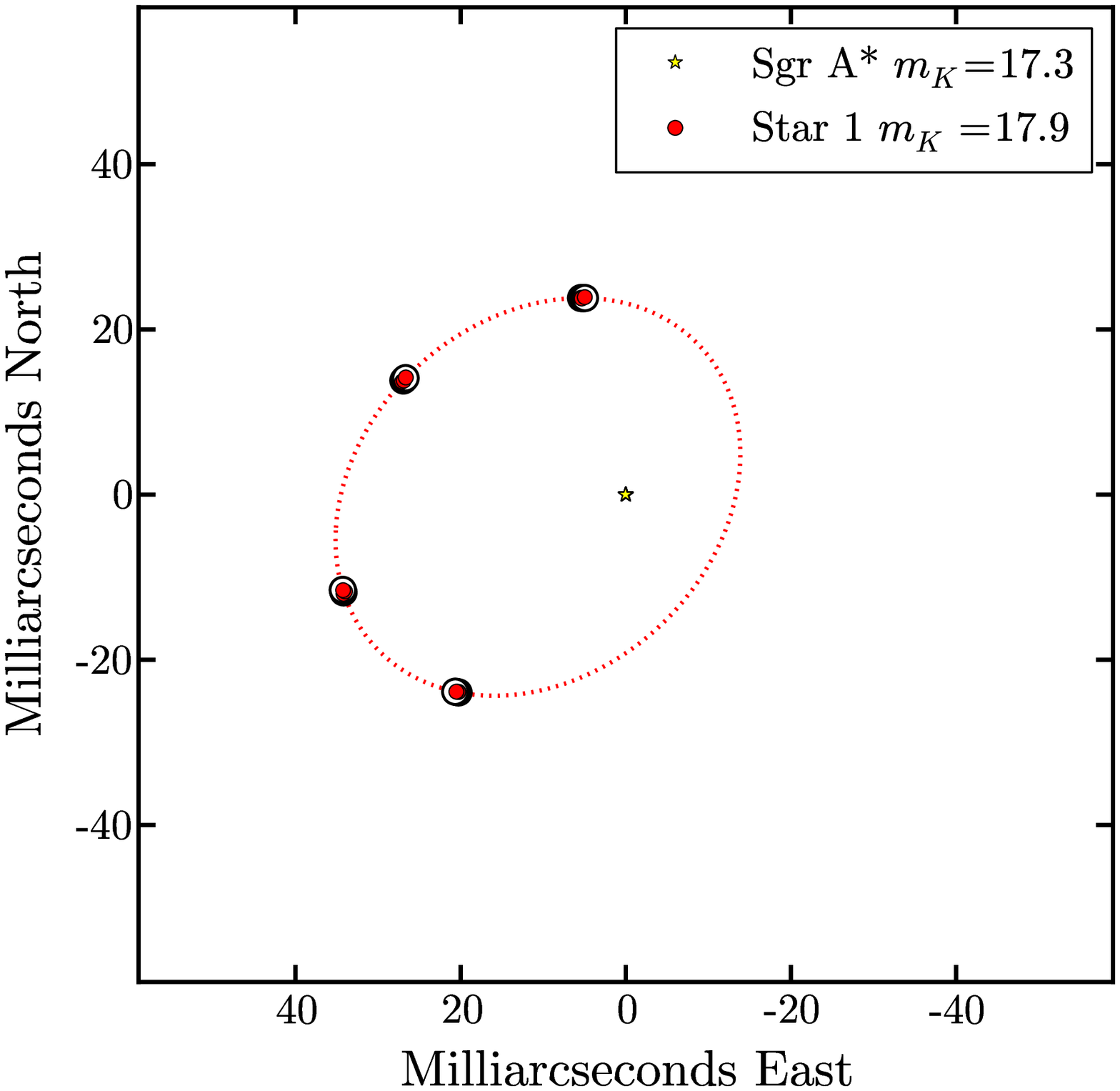}{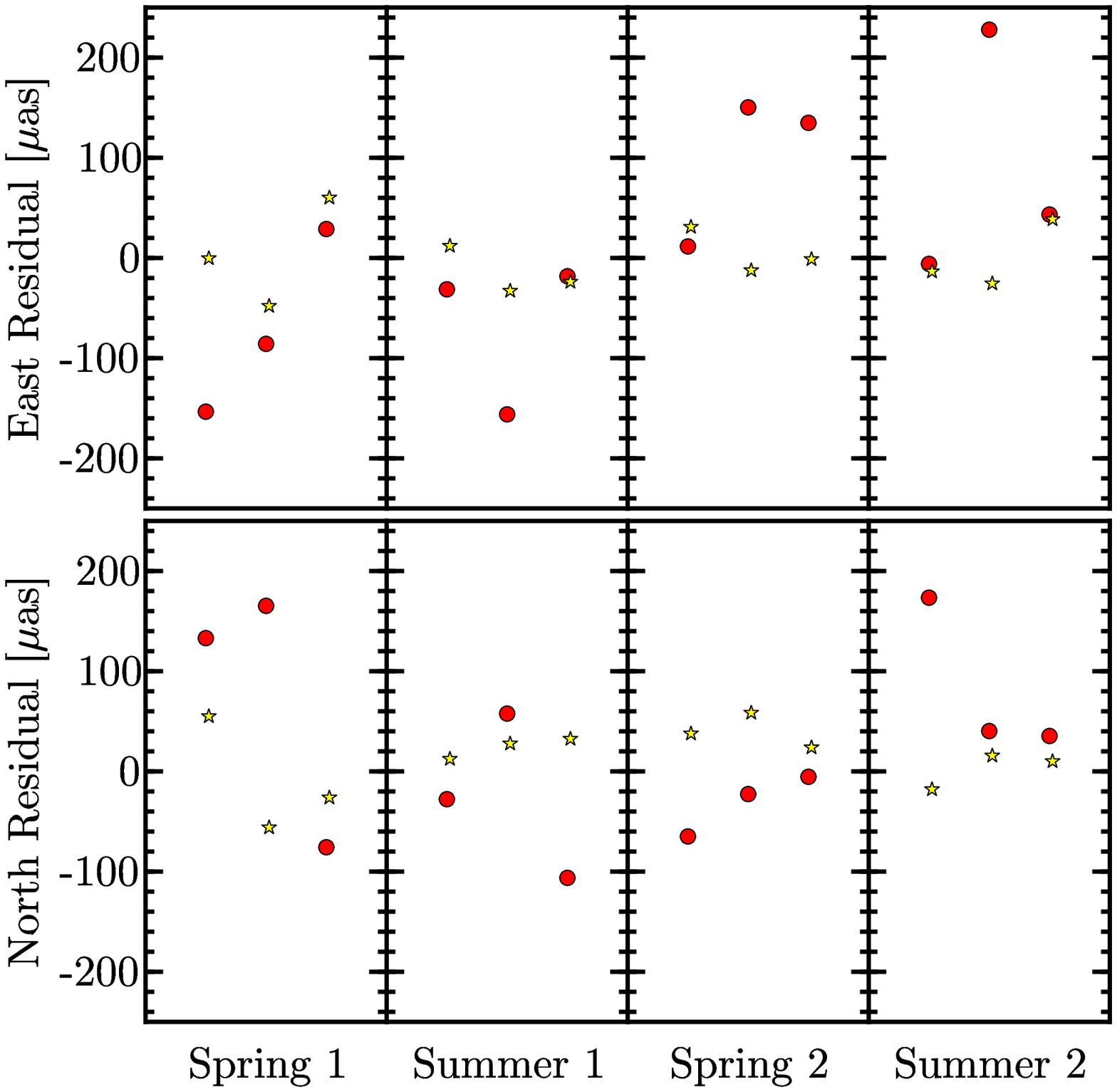}
\caption{
This plot is the same as Figure \ref{VLTField1Orbit} but for Field2. In this
case only the positions of Sgr A* and Star 1 are recovered. Error ellipses are
plotted for each point. White symbols indicate input locations.
\label{VLTField2Orbit}}
\end{center}\end{figure}

Figure \ref{VLTField2Orbit} shows our results fitting to simulated GRAVITY data
of Field2. We plot only the fitted positions for Sgr A* and Star 1 because no
other sources were confidently recovered. At input magnitudes
of $m_K=17.3$ and $m_K=17.9$, our astrometric residuals for Sgr A* and Star
1 are $\sim50\mu$as and $\sim150\mu$as respectively.

To evaluate the limiting factors in our GRAVITY observations, we run diagnostic
tests similar to those we perform for our ASTRA simulations. We start by
calculating upper limits to the signal-to-noise ratio of each source. Each
fringe generated by GRAVITY will have less photons than the corresponding
observation with ASTRA at the Keck Interferometer. First, because GRAVITY combines the
light from four telescopes between six baselines, only $\frac{2}{3}$ of the
flux incident on each aperture is available for combination. Second, the
individual apertures at the VLTI are smaller than the apertures at Keck.
Finally, the transmission of the GRAVITY instrument is expected to be less than
the transmission of ASTRA at Keck. 

Table \ref{VLTSNField1} shows upper limits to the signal-to-noise ratio in our
GRAVITY simulations calculated for each source in Field1. We see that for the
brightest sources, where injection fluctuations dominate the noise at Keck,
GRAVITY will provide a higher signal-to-noise ratio because with 6 baselines,
the effect of injection fluctuations averages down.  For fainter sources,
GRAVITY will provide a lower signal-to-noise because, as mentioned above, the
light is split more ways and the transmission is lower. Table \ref{VLTSNField1}
indicates, even with upper limits to the signal to noise, that Stars 3 and
4 will not be detectable in our simulations.
\begin{deluxetable}{ccccccc}
\tabletypesize{\scriptsize}
\tablewidth{0pt}
\tablecaption{VLTI simulated signal-to-noise ratios for each spectral channel for Field1}
\tablehead{
    \colhead{Spectral Channel} &
    \colhead{Sgr A* (k=16.3) } & 
    \colhead{Star 1 (k=16.9) } & 
    \colhead{Star 2 (k=18.8) } & 
    \colhead{Star 3 (k=20.3) } & 
    \colhead{Star 4 (k=20.6) }  
    }
\startdata
2.00 microns& 101 & 74 & 12 & 3 & 2 \\     
2.09 microns& 109 & 69 & 10 & 3 & 1 \\     
2.18 microns& 102 & 58 &  9 & 4 & 3 \\     
2.28 microns&  92 & 50 &  9 & 3 & 1 \\     
2.38 microns&  82 & 40 &  6 & 1 & 1 \\     
\enddata
\tablecomments{
This table shows upper limits to the simulated signal-to-noise ratio of each
source in Field1 provided by our model of the VLTI GRAVITY instrument. 
\label{VLTSNField1} }
\end{deluxetable}

Table \ref{VLTSNField2} shows upper limits to the signal-to-noise ratios for
the sources in Field2. These ratios indicate that no detectable signal is
present from Stars 2, 3, and 4 in Field2.
\begin{deluxetable}{ccccccc}
\tabletypesize{\scriptsize}
\tablewidth{0pt}
\tablecaption{VLTI simulated signal-to-noise ratios for each spectral channel for Field2}
\tablehead{
    \colhead{Spectral Channel} &
    \colhead{Sgr A* (k=17.3) } & 
    \colhead{Star 1 (k=17.9) } & 
    \colhead{Star 2 (k=19.8) } & 
    \colhead{Star 3 (k=21.3) } & 
    \colhead{Star 4 (k=21.6) }  
    }
\startdata
2.00 microns& 43 & 30 & 4 & 2 & 0 \\      
2.09 microns& 43 & 30 & 4 & 1 & 0 \\       
2.18 microns& 42 & 26 & 3 & 0 & 0 \\       
2.28 microns& 36 & 18 & 1 & 1 & 1 \\       
2.38 microns& 33 & 16 & 1 & 0 & 1 \\       
\enddata
\tablecomments{
This table shows simulated upper limits to the signal-to-noise ratio of each
source in Field2 provided by our model of the VLTI GRAVITY instrument. 
\label{VLTSNField2} }
\end{deluxetable}

\begin{figure}[h!] 
\begin{center} 
\epsscale{0.5}
\plotone{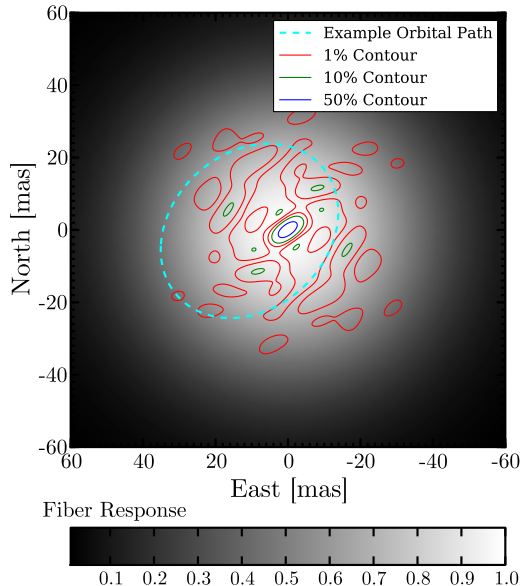} 
\caption{This figure is the same as Figure \ref{KeckBeamOrbit} but shows the
GRAVITY PSF (red, green, and blue contours at 1\%, 10\%, and 50\% respectively).
In addition, the fiber response function (grayscale) and an example stellar
orbit (cyan dashed line) are also plotted.\label{VLTBeamOrbit}} \end{center}\end{figure}

In Figure \ref{VLTBeamOrbit}, we see that our ability to accurately detect and
track stars in the vicinity of Sgr A* will depend on the exact location of the
star.  For example, within $~25$ mas of Sgr A*, where the optical fiber
transmits light most strongly, the PSF is at or above the 1\% level. Thus, as
in the Keck case discussed above, there is a tug-of-war of considerations
affecting the detectability of a source in the vicinity of Sgr A*. Faint
sources are more easily detected outside of the 1\% contours of the PSF. Because of
the fiber attenuation function, when sources are far from the center of the
field their flux level is likely to drop below the detection limit ($m_K\sim19$
in six hours).

\begin{figure}[h!] \begin{center}
\epsscale{0.8}
\plotone{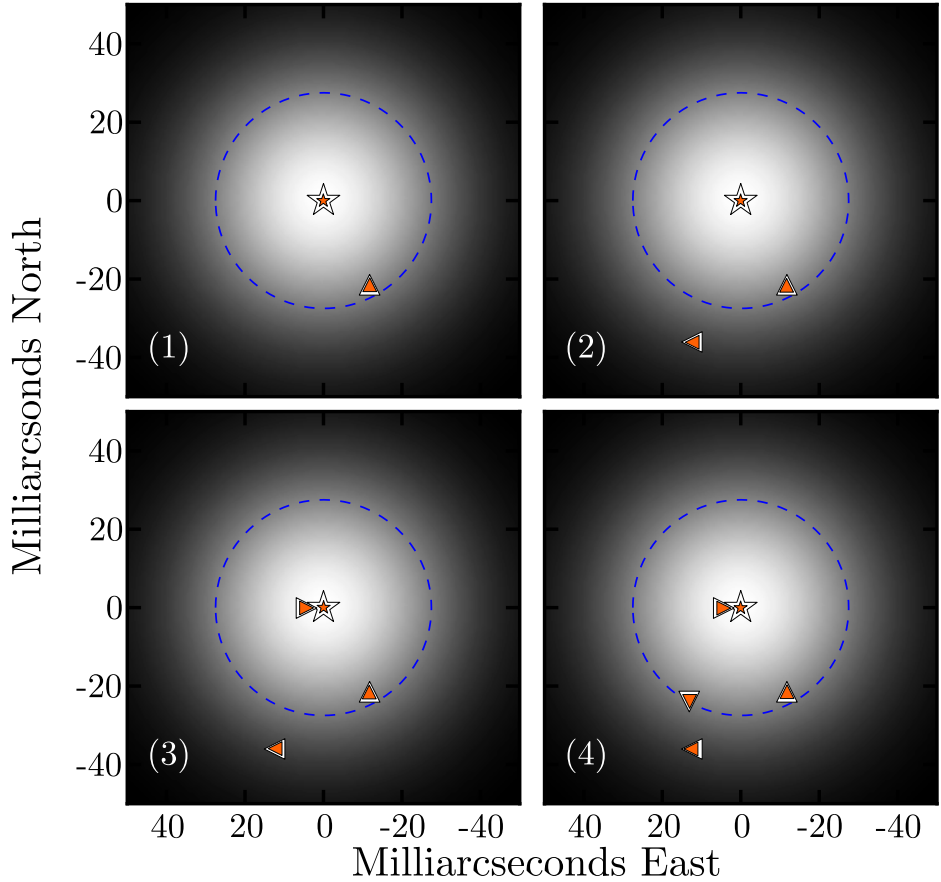}
\caption{ 
Our recovered positions (orange symbols) are plotted showing our performance
recovering source positions from GRAVITY data when more and more sources are
present. The input star field for each panel includes Sgr A* at the origin with
$m_K=16.3$ (star symbol). In addition to Sgr A* each star field also includes
from 1 to 4 $m_K=17$ stars (isosceles triangles with vertices pointing up,
right, left, and down corresponding to the first, second, third and fourth
added star respectively). Bootstrap error ellipses are smaller than the
plotting symbols. Also plotted is the Gaussian fiber response function
(grayscale) and the 50\% contour of this function (blue dashed line). Source
fluxes are attenuated by this function before detection.
\label{VLTconfusionFig}} \end{center}\end{figure}

While more beam splits reduce the number of photons combined between each pair
of telescopes at VLTI, the trade off is significantly increased uv-coverage. In
fact, the confusion limit for GRAVITY will be better than for ASTRA at Keck. In
Figure \ref{VLTconfusionFig}, as in Figure \ref{confusionFig} for the ASTRA,
we attempt to isolate the contribution of source crowding to the confusion
noise by observing star fields with more and more equal magnitude stars. As
in Figure \ref{confusionFig}, we start with Sgr A* at $m_K=16.3$ and one star
with $m_K=17$. We then add one $m_K=17$ star at a time until a total of four
stars are in the field. Since the VLTI provides good uv-coverage of the
Galactic Center, precise astrometry on even five bright sources within $\sim50$
mas of Sgr A* is possible. 

\begin{figure}[h!]
\begin{center} 
\epsscale{0.5} 
\plotone{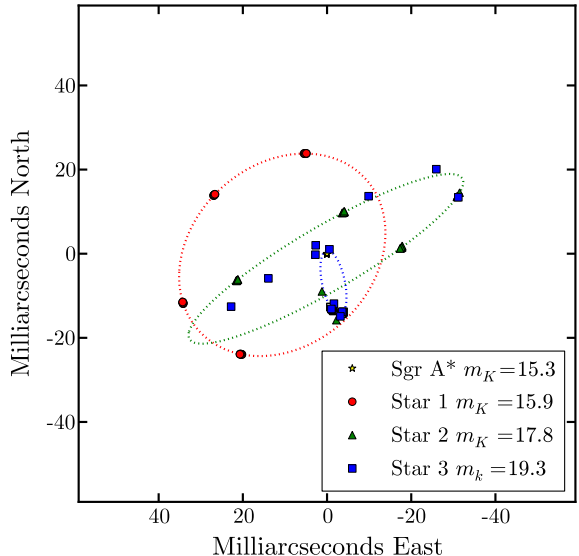}
\caption{Here we show our fits to 12 nights of simulated GRAVITY data for
Field0. Some difficulty in deducing the position of Star 2 occurs when it is
farthest off axis. During these nights, the flux of Star 2 is most attenuated
by the fiber response.  Since Star 3 is located close to on-axis for all
epochs, it appears as similarly bright to Star 2 during these nights. For two
nights during this run, our fitter has interchanged Star 2 and Star 3.
\label{VLTField0}} \end{center}\end{figure}

Since Table \ref{VLTSNField1} indicates that GRAVITY will be unable to recover
Stars 3 and 4 due to inadequate signal-to-noise, we also tested the performance on
a brighter star field labeled Field0. Field0 is identical to Field1 but with the flux
of each source increased by one magnitude. Our fits are plotted in Figure
\ref{VLTField0}. We are again able to recover the positions Sgr A*, Star 1 and
Star 2 for three out of four runs. During the run when Star 2 is farthest off
axis we again run into some difficulty, because our fitter interchanges Star
2 and Star 3.  This is due to the effects of the fiber response function which
attenuates the flux from Star 2 most during this run while Star 3 remains at
a nearly constant brightness near the center of the field. The interchange is
a result of our iterative fitting algorithm. This interchange does not affect
the results of multi-epoch observations which can track the sources and
identify the switch.

Star 3 in Field0 is recovered only less than half of the time, indicating that
the star is only marginally detectable in the data and suggesting a sensitivity
limit around $m_K\sim19$.  Even so, these results imply the potential to detect
and monitor several moderately bright sources within $\sim50$ mas of Sgr A*
should they exist.  

\section{Discussion} 

In Section \ref{ResultsSec}, our Keck Interferometer ASTRA simulations show the
ability to detect and track a stellar source on an orbit within $\sim50$ mas of
Sgr A* with a single baseline interferometer. This performance depends on the
source contrast and position angle with respect to Sgr A*. However, we
demonstrate the potential to detect and track an $m_K\sim18$ star when Sgr A*
is at $m_K=17.3$ and the star and Sgr A* are well separated along the baseline
direction (left panel of Figure \ref{Field2Select}).  We show that
$\sim150~\mu$as astrometry is possible along the baseline direction and $\sim3$
mas precision is possible in the transverse direction during unconfused epochs.
Our precision improves if the sources are brighter. Since a single
baseline interferometer produces an extended PSF, confusion still affects our
ability to accurately detect and track sources when their sidelobes overlap.

While at first glance ASTRA seems quite limited compared to GRAVITY in its
ability to detect and track stars within $\sim50$ mas of Sgr A*, we show that
the single baseline instrument could significantly contribute to the Galactic
Center science case. Specifically, we show that multi-epoch observations have
the ability to distinguish whether the region contains no bright sources, one
or two bright sources, or several bright sources (see Figures
\ref{Field1Select}, \ref{Field2Select}, and \ref{confusionFig}). Since the
source content in the region is truly unknown, any additional information about
the stellar density near the Galactic Center would be quite valuable. New
information could inform, for example, dynamical theories of the nuclear
cluster which must explain the positions of the stars.  Further, the higher
throughput, larger apertures, and fewer beam splits provide a larger signal in
the ASTRA fringes (compare for example Tables \ref{SNField2} and
\ref{VLTSNField2}). While confusion noise will constrain ASTRA observations, if
no star brighter than $m_K\sim19$ exists in the field ASTRA may have an
advantage in making a detection.  Because ASTRA is currently capable of making
these observations the potential exists to provide the community with some
constraints before GRAVITY comes online at the VLTI and before ASTRA operations
cease at Keck. 

While long-term operations of ASTRA are not currently planned, we note that if
ASTRA and GRAVITY observations could be obtained contemportaneously, some
importvement in the recovery of faint sources may be possible. In simulations
where we combined ASTRA and GRAVITY data (using the assumptions in Table
\ref{VisParams} for each instrument), we found that we can recover the
$m_K=19.8$ source in some epochs where it is not recovered using either ASTRA
or GRAVITY data alone.

Our GRAVITY simulations show that that instrument will attain a lower confusion
limit than ASTRA. This lower confusion in GRAVITY observations
is due to the increased uv-coverage provided by the VLTI array. This makes it possible
to detect and monitor multiple sources in the field. We demonstrate the
potential for $\sim10~\mu$as precision astrometry on Sgr A* at $m_K=16.3$ and
$\sim100~\mu$as precision astrometry on sources as faint as $m_K=18.8$ in six
hours of observing our simulated star fields.  However, we show that the
decreased throughput at GRAVITY and the larger number of beam splits required
to create six baselines will impose a detection limit at GRAVITY which will
make it difficult to detect sources at $m_K\gtrsim19$.  In fact, we show that
for one simulated three-night run using GRAVITY, we are unable to recover the
position of the $m_K=18.8$ source.  While this reflects contrast and fiber
function issues to some extent, it shows that in real observations GRAVITY may
struggle to detect sources at this brightness level.

GRAVITY's ability to recover precise astrometry for multiple sources within
$\sim50$ mas from Sgr A* suggests it should be able to constrain the shape of
an extended mass distribution at the Galactic Center\citep{Rubilar2001,
Weinberg05}. Any model of the central structure must include the mass of the
black hole and the mass and radial profile of an extended distribution of
matter. To constrain these parameters and to break the first-order degeneracy
between the retrograde precession due to the extended matter and the prograde
precession attributable to General Relativity, multiple stars with distinct
angular momenta will be needed \citep{Rubilar2001, Weinberg05}. Since the
astrometric signal of orbital precession increases linearly with the number of
revolutions, monitoring stars on short-period orbits within $\sim50$ mas is
preferred, since a larger signal can be detected in shorter time.

Our GRAVITY simulations also demonstrate the astrometric precision needed to
detect relativistic effects on stellar orbits. Our simulated performance of
$\sim100~\mu$as suggests that low order effects of relativity, such as the
prograde precession, will be detectable \citep{Weinberg05}. However, higher
order relativistic effects, such as detecting the influence of the black hole
spin on stellar orbits, will be more difficult, requiring measurements more
precise than those demonstrated here\citep{Weinberg05, Merritt2011}.

A recent paper by \citet{Vincent11} modeled the imaging mode astrometric
performance of GRAVITY, applying the CLEAN algorithm to images formed using the
interferometric visibilities. In that paper, the authors mainly investigate
the astrometric precision attainable on Sgr A* when it is very bright. They
compare their performance after a whole night of observing to individual 100
second exposures. They show that $\sim40~\mu$as precision is attainable on
Sgr A* in 100 seconds when it is very bright and the field is simple. Our
simulations show that astrometric precision on the order of the angular extent
of the inner-most stable circular orbit of the black hole ($\sim30~\mu$as) is
attainable on Sgr A* even in the midst of our more complicated star field.
However, $\sim10~\mu$as precision is attained after 6 hours of
observing; time resolved astrometry in our fields will necessarily be less
precise. Not only are shorter observations less sensitive to sources in the
field, but with less time spent on-source the uv-coverage is reduced. Both of
these effects combine to degrade the astrometric precision by decreasing the
signal-to-noise ratio and increasing the confusion. We demonstrate that as
astrometry on Sgr A* becomes more difficult due to confusion with bright
stars in the small field, astrometry on those bright stars becomes easier.
Thus GRAVITY should provide some traction on investigating General Relativistic
effects, either through observations of Sgr A* itself given a faint star field
or by tracking stars in the vicinity of Sgr A* given a brighter star field.

Although our simulations did not include Sgr A* variability explicitly,
variability could provide an interesting paradigm for making these
observations. During high states, we will be able to conduct precise astrometry
of Sgr A*, anchoring our field. During low states, the decreased contrast will
provide an opportunity to probe for fainter stellar sources in the region. This
back-and-forth approach could be harnessed to precisely monitor faint sources.
To demonstrate these effects, we ran our simulator with Sgr A* set to very low
brightness but with the star field magnitudes kept constant. During these runs,
we were able to recover stellar positions more easily, since confusion with Sgr
A* was reduced. On the other hand, a very bright Sgr A* is easily detected with
a high level of precision. The timescales of Sgr A* variability are conducive to
seeing both high and low states while observing. Flares are observed on
timescales of $\sim10-100$ minutes and Sgr A* often changes flux by more than
one magnitude.

\section{Summary and Future Work} 
Our simulations demonstrate that ASTRA and GRAVITY will be able to provide
different insights into the star field near Sgr A*. The Keck instrument will
excel if the inner $\sim50$ mas is a simple field, with a steep luminosity
function including at least one relatively bright star. We demonstrate the
ability to recover a source with $m_K=17.9$ in a field with other similarly
bright sources and we show in Table \ref{SNField2} that a source as faint as
$m_K=19.8$ might be detected if the star field is faint and the visibility
amplitude is high. Multi-epoch observations will be necessary to mitigate
source confusion as the astrometry will be most precise when the star orbits
through position angles where the astrometric offset along the projected
baseline direction is large.  GRAVITY's sensitivity to sources will not depend
strongly on orbital phase as is the case with ASTRA since GRAVITY provides
a more symmetric PSF. Moreover, GRAVITY will be better able to track orbits if
the stellar field has a shallower luminosity function, as it is not as affected
by confusion noise and because it will have difficulty detecting sources
fainter than $m_K\sim19$.

The minimal uv-coverage provided by the single baseline of the Keck
Interferometer, which is furthermore situated in the northern hemisphere where
Sgr A* transits low, is not insufficient for providing valuable information for
scientific advance. In fact, we demonstrate the ability to detect and monitor
stars when there is sufficient astrometric offset along the baseline direction.
These results could be extended to infer the performance of a single baseline
of the VLTI. Given the technical and practical challenges of using all four VLT
apertures to create six baselines, it is important to consider that even
a reduced array at VLTI could make important contributions to Galactic Center
science. Because the VLTI is situated in the southern hemisphere where Sgr A*
transits high, even a single baseline of the VLTI would provide much more
uv-coverage of Sgr A* than Keck. If no bright stars are detected in the region,
then a reduced array could be used to provide more photons in each fringe,
increasing the sensitivity of the interferometer at the expense of a more
extended PSF, which would even so be less extended than the Keck Interferometer
PSF we show above.  

Moving forward, further GRAVITY simulations incorporating a variable Sgr A*
and stars on post-Newtonian orbits will be useful in the interim before
that instrument comes online. Such simulations will aid in predicting the
challenges of characterizing the gravitational potential at the Galactic Center
with stellar orbits and in creating the necessary analysis tools which will be
needed for fitting the complicated orbits which are expected to be observed.

Finally, some interferometric observations of Galactic Center sources have been
made at Keck and the VLTI \citep[e.g.,][]{Pott2008A}. \citet{Pott2008B}
observed IRS 7 with the VLTI and showed it is suitable for use as a phase
reference source; similar observations have been made with the Keck
Interferometer. Dual field phase referencing has been demonstrated on-sky with
ASTRA (Woillez et al. in prep), and the instrument is poised to
observe the field around Sgr A*. An obvious next step is to actually observe
Sgr A* and to search for real sources.

\section{Aknowledgements} 
We would like to thank and aknowledge the Keck Interferometer ASTRA team for
their work on the instrument and for valuable input in designing our simulator.
JAE gratefully acknowledges support from an Alfred P. Sloan Research
Fellowship.

\end{document}